\documentclass[%
 reprint,
 bibnotes,
 amsmath,amssymb,
 aps,
]{revtex4-2}

\usepackage{graphicx}
\usepackage{dcolumn}
\usepackage{bm}

\usepackage[hidelinks,colorlinks=true]{hyperref} 
\usepackage{cleveref}
\usepackage{enumitem}
\usepackage{xspace}
\usepackage{siunitx}
\usepackage{lineno}
\usepackage{multirow}
\usepackage{makecell}
\usepackage[dvipsnames]{xcolor}
\definecolor{darkred}{rgb}{0.5,0,0}
\definecolor{darkblue}{rgb}{0,0,0.5}
\definecolor{firebrick}{rgb}{0.75,0.125,0.125}
\definecolor{darkgreen}{rgb}{0,0.5,0}
\hypersetup{urlcolor=darkblue,
	    citecolor=darkgreen,
	    linkcolor=firebrick}

\newcommand{\nue}{\ensuremath{\nu_{e}}\xspace}
\newcommand{\numu}{\ensuremath{\nu_{\mu}}\xspace}
\newcommand{\nutau}{\ensuremath{\nu_{\tau}}\xspace}

\newcommand{\obs}[1]{\ensuremath{{#1}^{\rm obs}}}

\newcommand{\mathintitle}[1]{\texorpdfstring{$\mathbf{#1}$}{null}\xspace}

\newcommand{\ie}{{i.e.}}

\newcommand{\eg}{{e.g.}}

\newcommand{\eq}{Eq.}
\newcommand{\fig}{Fig.}
\newcommand{\Refe}{Ref.}
\newcommand{\Refes}{Refs.}
\newcommand{\equ}[1]{\eq~(\ref{equ:#1})}
\newcommand{\figu}[1]{\fig~\ref{fig:#1}}

\begin{document}

\title{The flavor composition of ultra-high-energy cosmic neutrinos: measurement forecasts for in-ice radio-based EeV neutrino telescopes}

\author{Alan Coleman}
\email{alan.coleman@physics.uu.se}
\author{Oscar Ericsson}
\affiliation{Department of Physics and Astronomy, Uppsala University, Uppsala, SE-752 37, Sweden}

\author{Mauricio Bustamante}
\email{mbustamante@nbi.ku.dk}
\affiliation{Niels Bohr International Academy, Niels Bohr Institute, University of Copenhagen, DK-2100 Copenhagen, Denmark}

\author{Christian Glaser}
\email{christian.glaser@physics.uu.se}
\affiliation{Department of Physics and Astronomy, Uppsala University, Uppsala, SE-752 37, Sweden}

\date{\today}

\begin{abstract}
In-ice radio-detection is a promising technique to discover and characterize ultra-high-energy (UHE) neutrinos, with energies above $\SI{e17}{eV}$, adopted by present---ARA, ARIANNA, and \hbox{RNO-G}---and planned ---IceCube-Gen2.  So far, their ability to measure neutrino flavor had remained unexplored.  We show and quantify how the neutrino flavor can be measured with in-ice radio detectors using two complementary detection channels.  The first channel, sensitive to $\nu_e$, identifies them via their charged-current interactions, whose radio emission is elongated in time due to the Landau-Pomeranchuk-Migdal effect.  The second channel, sensitive to $\nu_\mu$ and $\nu_\tau$, identifies events made up of multiple showers generated by the muons and taus they generate.  We show this in state-of-the-art forecasts geared at IceCube-Gen2, for representative choices of the UHE neutrino flux.  This newfound sensitivity could allow us to infer the UHE neutrino flavor composition at their sources---and thus the neutrino production mechanism---and to probe UHE neutrino physics.
\end{abstract}

\maketitle

\section{Introduction}
\label{sec:intro}

Ultra-high-energy (UHE) cosmic neutrinos, with energies larger than \SI{e17}{eV}, hold vast potential to probe astrophysics and fundamental physics at the highest energies.  They were first predicted more than fifty years ago~\cite{Berezinsky:1969erk}, as byproducts of the interaction of ultra-high-energy cosmic rays (UHECRs) with the cosmic microwave background~\cite{Greisen:1966jv, Zatsepin:1966jv}.  However, because their flux is low, they have eluded detection so far~\cite{IceCube:2018fhm, PierreAuger:2019ens}.  Soon, though, a new generation of large-scale UHE neutrino telescopes~
\cite{MammenAbraham:2022xoc, Ackermann:2022rqc, Guepin:2022qpl, Barwick:2022vqt}, presently in different stages of planning, construction, and prototyping, could finally discover them, realizing long-awaited opportunities. 

\begin{figure}[t!]
 \centering
 \includegraphics[width=\columnwidth]{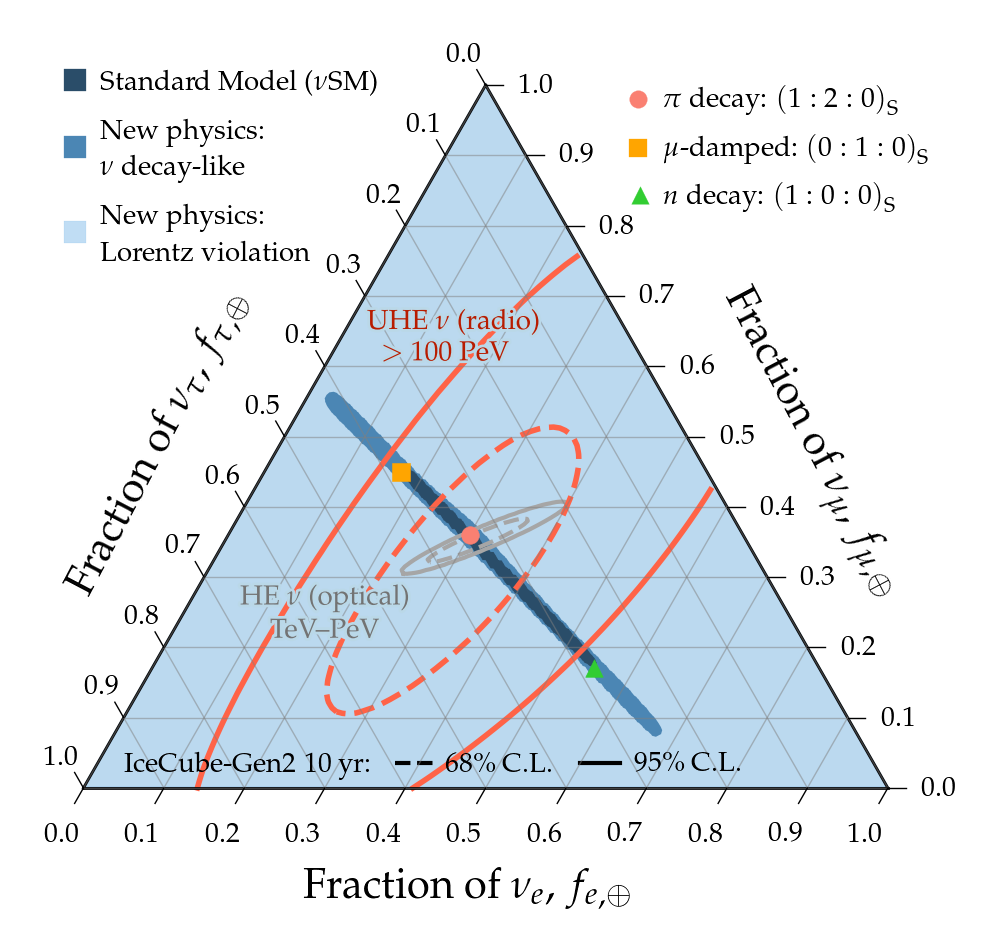}
 \caption{\textbf{\textit{Measurement forecasts of the flavor composition of the diffuse flux of ultra-high-energy (UHE, $\gtrsim 100$~PeV) cosmic neutrinos.}}  We show the first forecasts of flavor measurements via neutrino radio-detection, geared at the planned radio array of IceCube-Gen2~\cite{IceCube-Gen2:2020qha}. For comparison, we show also forecasts for high-energy (HE, TeV--PeV) neutrinos, using optical detection in IceCube-Gen2~\cite{IceCube-Gen2:2023rds}.  In this figure, for illustration, we assume a large benchmark UHE neutrino flux (\figu{fluxes}) produced in pion decays.  Our results are based on state-of-the-art simulations of in-ice neutrino radio-detection.  The predicted regions of allowed flavor composition at Earth~\cite{Bustamante:2015waa}, under standard oscillations and new physics, use narrow ranges of the neutrino mixing parameters foreseen for 2040~\cite{Song:2020nfh}. \textit{Measuring the UHE flavor composition will further our insight into neutrino physics and astrophysics.} See also Fig.~\ref{fig:flavour_energy} for an energy-dependent flavor measurement.}
 \vspace*{-0.7cm}
 \label{fig:ternary_he_vs_uhe}
\end{figure}

Measurement of the flavor composition of UHE neutrinos---\ie, the proportion of \nue, \numu, and \nutau in their flux---represents one such opportunity, one with the power to deliver remarkably versatile insight.  
For astrophysics, the flavor composition reflects the physical conditions present inside the astrophysical neutrino sources and so may hint at their identity, which so far remains unknown~\cite{Rachen:1998fd, Athar:2000yw, Crocker:2001zs, Barenboim:2003jm, Beacom:2003nh, Beacom:2004jb, Kashti:2005qa, Mena:2006eq, Kachelriess:2006ksy, Lipari:2007su, Esmaili:2009dz, Choubey:2009jq, Hummer:2010ai, Winter:2013cla, Palladino:2015zua, Bustamante:2015waa, Biehl:2016psj, Bustamante:2019sdb, Ackermann:2019ows, Bustamante:2020bxp, Song:2020nfh, Bhattacharya:2023mmp, Telalovic:2023tcb, Dev:2023znd}.  For fundamental physics, it probes the standard paradigm of neutrino oscillations and may reveal physics beyond the Standard Model~\cite{Beacom:2002vi, Barenboim:2003jm, Beacom:2003nh, Beacom:2003eu, Beacom:2003zg, Serpico:2005bs, Mena:2006eq, Lipari:2007su, Pakvasa:2007dc, Esmaili:2009dz, Choubey:2009jq, Esmaili:2009fk, Bhattacharya:2009tx, Bhattacharya:2010xj, Bustamante:2010nq, Mehta:2011qb, Baerwald:2012kc, Fu:2012zr, Pakvasa:2012db, Chatterjee:2013tza, Xu:2014via, Aeikens:2014yga, Arguelles:2015dca, Bustamante:2015waa, Pagliaroli:2015rca, Shoemaker:2015qul, deSalas:2016svi, Gonzalez-Garcia:2016gpq, Bustamante:2016ciw, Rasmussen:2017ert, Dey:2017ede, Bustamante:2018mzu, Farzan:2018pnk, Ahlers:2018yom, Brdar:2018tce, Palladino:2019pid, Ackermann:2019cxh, Arguelles:2019rbn, Ahlers:2020miq, Karmakar:2020yzn, Fiorillo:2020gsb, Song:2020nfh, Arguelles:2022tki, MammenAbraham:2022xoc, Telalovic:2023tcb}. 
At TeV--PeV neutrino energies, the IceCube neutrino telescope regularly measures the flavor composition of high-energy cosmic neutrinos~\cite{Mena:2014sja, Palomares-Ruiz:2015mka, IceCube:2015rro, Palladino:2015vna, IceCube:2015gsk, Vincent:2016nut, IceCube:2020fpi}.
However, at ultra-high energies, the measurement of the neutrino flavor composition is largely unexplored. In the following, we address this shortcoming.

In-ice radio-detection~\cite{Barwick:2022vqt} is a promising and mature technique for measuring UHE neutrinos.  Neutrino detectors that adopt it consist of compact detector stations made up of a handful of antennas,  deployed in polar ice sheets at shallow depths.  They search for nanosecond-long radio flashes, known as Askaryan radiation~\cite{Askaryan:1961pfb}, generated by neutrino interactions in the ice.  This allows for the cost-efficient instrumentation of huge volumes needed to detect the low flux of UHE neutrinos. The technical feasibility of the technique was demonstrated in the pathfinder arrays RICE, ARA, and ARIANNA at the South Pole and on the Ross Ice Shelf in Antarctica~\cite{RICE:2001ayk,Anker:2019rzo, ARA:2019wcf,ARIANNA:2020zrg,Arianna:2021lnr,ARA:2022rwq}. 
Currently, the Radio Neutrino Observatory in Greenland (RNO-G) is under construction~\cite{RNO-G:2020rmc} which has the potential to measure the first UHE neutrino. At the same time, an order-of-magnitude more sensitive radio detector is in advanced planning stages as part of  IceCube-Gen2~\cite{IceCube-Gen2:2020qha,IceCube-Gen2:2021rkf,IceCube-Gen2-TDR}. Depending on the flux of UHE neutrinos, tens to hundreds of them could be observed over the lifetime of IceCube-Gen2~\cite{Valera:2022ylt, Fiorillo:2022ijt, Valera:2022wmu, Valera:2023ayh}, providing insight into the origin of UHECRs~\cite{Anchordoqui:2018qom, AlvesBatista:2019tlv}.

Here, for the first time, we forecast the ability to measure the flavor composition of UHE neutrinos at Earth using in-ice radio-detection in a large array like IceCube-Gen2. This work complements the forecasts of the other main science objectives of discovering and characterizing the UHE neutrino flux \cite{Valera:2022wmu,Valera:2023ayh},  identifying the sources of UHE neutrinos \cite{Fiorillo:2022ijt}, and measuring the UHE neutrino-nucleon cross section~\cite{Valera:2022ylt,Esteban:2022uuw},  
Our work builds up on the advances in modeling the signatures of UHE neutrinos in radio detectors through the {\tt NuRadioMC} code~\cite{Glaser:2019cws} and its extension to simulate secondary interactions of muons and taus~\cite{Garcia-Fernandez:2020dhb, Glaser:2021hfi}. 

We obtain sensitivity to the neutrino flavor through two complementary detection channels, which are described in detail in the next section. First, interference of multiple overlapping showers generated by \nue charged-current (CC) interactions alters the shape of the radio pulse, which can be detected with a deep neural network.  (Preliminary results on this were presented in \Refes~\cite{ericsson2021investigations,Stjarnholm:2021xpj}). Second, \numu and \nutau can be identified by detecting events containing at least two displaced in-ice showers, from their initial CC interaction or from the stochastic energy losses of the muon or tau that they generate.  From the combination of both detection channels, we devise, for the first time, sensitivity to the three neutrino flavors in in-ice radio-detection neutrino telescopes, opening up measurements of the flavor composition to the entire range of high-energy cosmic neutrino energies.

\Cref{fig:ternary_he_vs_uhe} summarizes our results, geared towards radio-detection in IceCube-Gen2, which we simulate in state-of-the-art detail.  For the high-flux assumption of UHE neutrino production, our method affords enough precision to distinguish between competing benchmark alternatives of neutrino production at the 95\%~confidence level (C.L.) in a baseline 10 years of observation.  This sensitivity also allows us to contrast between flavor transitions driven by standard oscillations and new physics at the highest energies, where differences may be more apparent.  Our methods and results are novel, timely, and do not require additional detector hardware or capabilities compared to those already envisioned.  They rely only on searching for distinct features in detected events.  

This paper is organized as follows.  In \Cref{sec:how_to_measure_flavor}, we describe the flavor-sensitive detection channels in in-ice radio detectors. In \Cref{sec:cnn}, we introduce a neural network to identify \nue CC interactions. In \Cref{sec:sensitivity}, we describe the generation of mock event samples and present our methods to measure  neutrino flavor.  In \Cref{sec:results}, we apply them to forecast the measurement of the UHE flavor composition in IceCube-Gen2. In \Cref{sec:discussion}, we discuss the impact of our results on UHE neutrino science. In \Cref{sec:summary}, we summarize and conclude.


\section{How to measure neutrino flavor with in-ice radio detectors}
\label{sec:how_to_measure_flavor}

Obtaining sensitivity to the neutrino flavor at ultra-high energies with in-ice radio detectors is challenging because the neutrino interactions of different flavors produce very similar radio signatures.

At these energies, the most likely interaction channel is neutrino-nucleon deep inelastic scattering ($\nu N$ DIS)~\cite{CTEQ:1993hwr, Conrad:1997ne, Formaggio:2012cpf, IceCube:2017roe, Bustamante:2017xuy, IceCube:2018pgc, IceCube:2020rnc}, where the neutrino scatters off of one of the constituent partons of a proton or neutron---most likely a quark---breaking up the nucleon, $N$, in the process. The interaction is either neutral-current (NC)---mediated by a $Z$ boson, \ie, $\nu_\alpha + N \to \nu_\alpha + X$, where $\alpha = e, \mu, \tau$, and $X$ represents final-state hadrons---or CC---mediated by a $W$ boson, \ie, $\nu_\alpha + N \to l_\alpha + X$. At these energies, the $\nu N$ DIS cross section is nearly equal for neutrinos of all flavor, and for neutrinos and anti-neutrinos.  Final-state charged particles initiate particle showers whose electromagnetic emission---radio, in our case---may be detectable. Flavor sensitivity stems exclusively from CC interactions.  Fortunately, because the CC $\nu N$ DIS cross section is roughly three times larger than the NC cross section, events from which to extract flavor sensitivity are not rare.

\begin{figure*}[t!]
 \centering
 \includegraphics[width=0.999\textwidth]{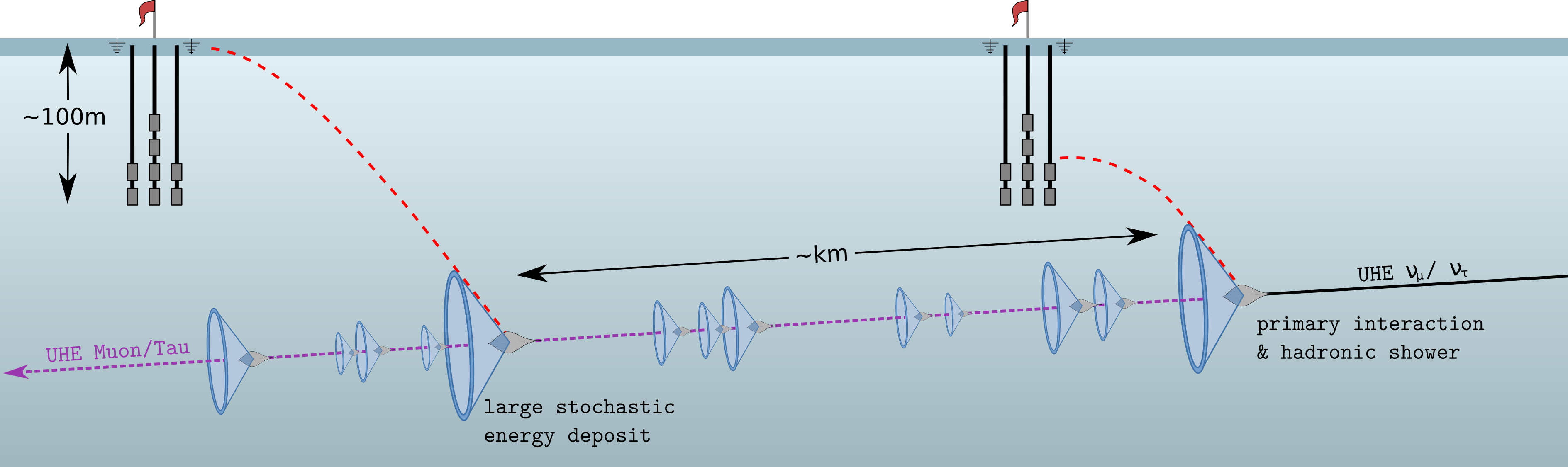}\\
 \caption{\textbf{\textit{Schematic of a charged-current interaction of a UHE \numu or \nutau and its in-ice radio-detection.}} The primary neutrino-nucleon interaction produces a hadronic shower. The final-state charged lepton, a muon or a tau, can travel several kilometers while producing sub-showers stochastically, generating Askaryan radiation which can be observed at multiple underground detector stations, resulting in a \emph{multi-shower} event.  The change of the index of refraction of radio with depth causes the trajectory of radio signals to bend on their way to the stations, an effect we account for in our simulations.  See \Cref{sec:how_to_measure_flavor-multi_shower} for details.  \textit{Detecting multi-shower events grants us access primarily to the UHE $\nu_\mu + \nu_\tau$ content (and also, indirectly, to the $\nu_e$ content); see \figu{observable_pairs}.}}
 \label{fig:multi_det_schematic}
\end{figure*}

In a NC interaction, only the final-state hadrons initiate showers. These \emph{hadronic showers} start with a high content of neutrons and pions, but as they evolve they quickly transfer their energy to electromagnetic showers (EM), made up mostly of electrons, positrons, and photons. The particle-content profile of these showers has a typical length of \SI{15}{m}, is smooth, increasing monotonically until the maximum is reached and then decreases monotonically, and exhibits little variation between showers; see \figu{multi_det_schematic}, bottom right, and the examples in \Refes~\cite{Glaser:2019cws, Barwick:2022vqt}. As they evolve, the EM showers develop a time-varying excess of negative charge that generates a nanosecond-scale impulsive radio signal---Askaryan emission~\cite{Askaryan:1961pfb}---which in-ice radio detectors target (\figu{multi_det_schematic}). Because in NC interactions the radio emission is solely from the final-state hadrons, all neutrino flavors produce the same event signature and are indistinguishable. 

In a CC interaction, both the final-state hadrons and lepton initiate showers. Like before, the hadronic shower triggered by the former does not provide flavor sensitivity.  The sensitivity stems from the additional Askaryan radiation generated by the final-state electron, muon, or tau---depending on whether the interacting neutrino is a \nue, \numu, or \nutau---whose  interactions yield different, distinguishable event signatures. We describe them below.  


\subsection{\mathintitle{\numu + \nutau} detection channel: radio emission from secondary muons and taus}
\label{sec:how_to_measure_flavor-multi_shower}

Figure~\ref{fig:multi_det_schematic} shows how, in addition to the hadronic shower located at the $\nu N$ interaction vertex, the muon or tau lepton generated in a CC \numu or \nutau interaction initiates, as it propagates through the ice, several high-energy secondary showers through stochastic energy losses and through decay, in the case of the tau~\cite{Garcia-Fernandez:2020dhb}. (The decay length of taus increases roughly linearly with energy and is already approximately~\SI{5}{km} at \SI{e17}{eV}~\cite{Glaser:2019cws}, longer than the station spacing considered for IceCube-Gen2. Therefore, tau decays play only a subdominant role at ultra-high energies; see \cite{Garcia-Fernandez:2020dhb} for details.) 

These showers are significantly displaced from the $\nu N$ interaction vertex, on the order of a kilometer. Each shower generates Askaryan emission in the same manner as the hadronic interaction at the $\nu N$ interaction vertex. This adds additional opportunities for detection, as shown by \Refe~\cite{Garcia-Fernandez:2020dhb}, increasing the effective detector volume for \numu and \nutau by 20--40\% compared to targeting only the radio emission from the $\nu N$ interaction vertex~\cite{Garcia-Fernandez:2020dhb, Glaser:2021hfi,ARA:2023vaf}. The fraction of events with detectable secondary showers for RNO-G~\cite{Garcia-Fernandez:2020dhb}, IceCube-Gen2~\cite{Glaser:2021hfi}, and ARA~\cite{Cummings:2023dbj, ARA:2023vaf} is similar.

In some events, neighboring detector stations could detect radio emission from the $\nu N$ interaction and from one of the secondary interactions, which would give a clear and background-free signature of \numu and \nutau~\cite{Garcia-Fernandez:2020dhb,Glaser:2021hfi}. The detection rate of these \textit{multi-shower} events depends on the neutrino energy, the layout of the detector array, the detector site, and trigger settings. The design parameter with the largest influence on the multi-shower detection rate is the station spacing: the closer neighboring stations are, the higher the probability of multi-shower detection. 

\begin{figure}[b!]
 \centering
 \includegraphics[width=\columnwidth]{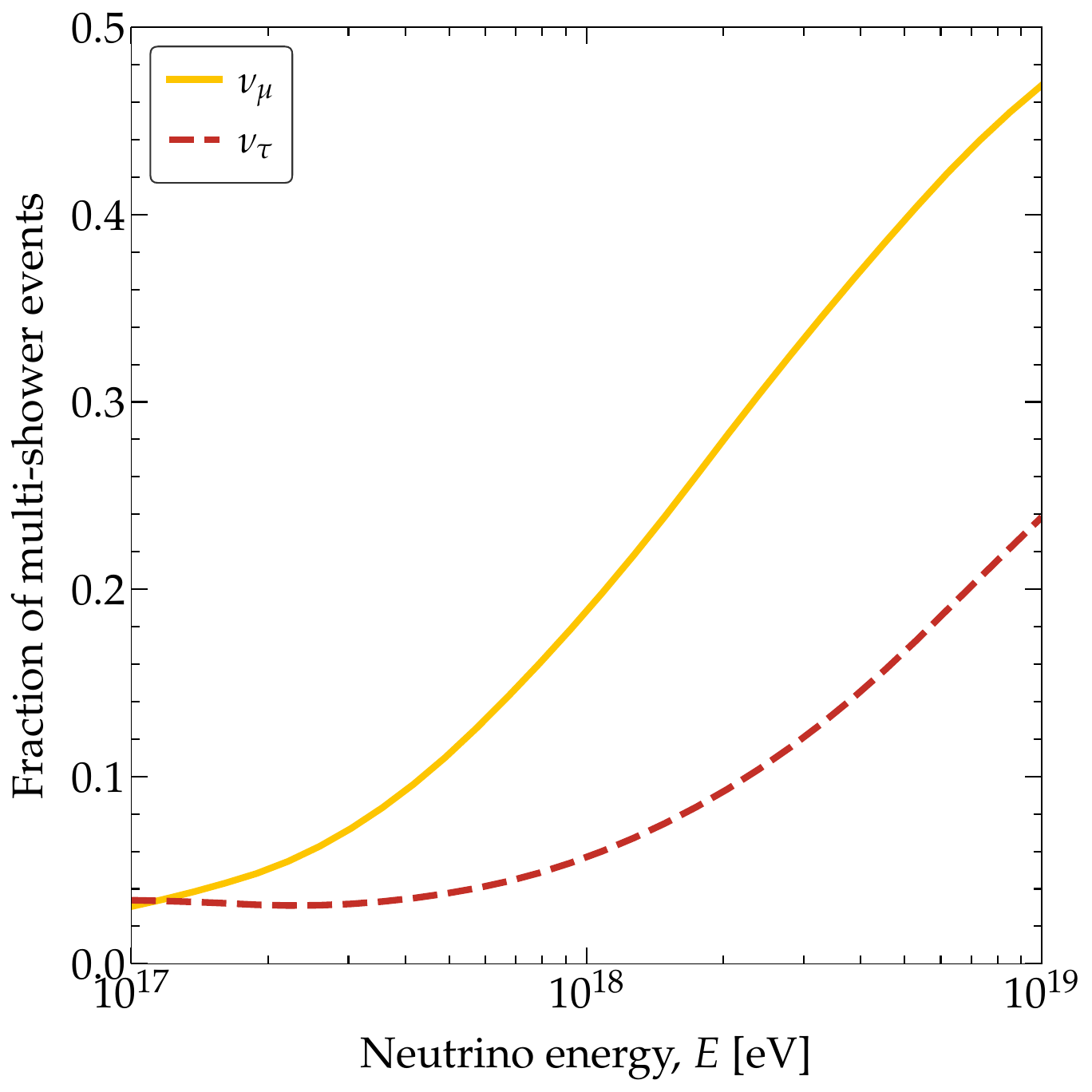}
 \caption{\textbf{\textit{Fraction of UHE \numu- and \nutau-initiated CC interactions that can be detected as multi-shower events.}}  We assume radio-detection in the radio array of IceCube-Gen2 with 2\,km spacing between detector stations. Data obtained from the detailed detector simulations of \Refe~\cite{Glaser:2021hfi}. See \Cref{sec:how_to_measure_flavor-multi_shower} for details.}
 \label{fig:fraction_multi_station}
\end{figure}

Figure~\ref{fig:fraction_multi_station} shows the expected fraction of CC interactions initiated by \numu and \nutau that are detected as multi-shower events in the radio array of IceCube-Gen2~\cite{Glaser:2021hfi}. Later, we use this fraction in our analysis. The fraction increases with energy, as the probability of large stochastic showers happening increases, and is significantly higher for \numu because, statistically, muons deposit larger fractions of their energy.
Yet, from a single detected multi-shower event, we do not foresee distinguishing if the interacting neutrino was a \numu {\it vs.}~a \nutau, but its detection alone rules out that it was a \nue.
In our analysis, the proportion of multi-shower events in a sample of detected events reflects the joint content of $\nu_\mu + \nu_\tau$ in the neutrino flux.  Nonetheless, since muons are more likely to trigger multiple stations, there is limited sensitivity separately to the relative abundance of \numu and \nutau (\Cref{sec:results}).

Assuming that the direction and position of the showers can be reconstructed sufficiently well, which is supported by current studies \cite{Glaser:2019rxw,ARIANNA:2021pzm,ARIANNA:2020zrg,Arianna:2021lnr,Aguilar:2021uzt,Glaser:2022lky,Plaisier:2023cxz,IceCube-Gen2:2023czw,IceCube-Gen2:2023wcw}, the only potential background to this detection channel are high-energy muons generated in cosmic-ray-induced air showers. However, their rate of detection by a single station is already low: we expect no more than a handful of events in ten years of operation of IceCube-Gen2~\cite{Pyras:2023crm}, clustered near the low-energy threshold of the experiment, between \SI{e16}{eV} and \SI{e17}{eV} (see also Figs.~3, 4, and 6 in \Refe~\cite{Valera:2022wmu}). Hence, the number of atmospheric muons capable of triggering showers in two neighboring stations is negligible; we ignore it in the following.


\subsection{\mathintitle{\nue} detection channel: identification of LPM-elongated showers}
\label{sec:how_to_measure_flavor-lpm}

\begin{figure*}[tb]
 \centering
 \includegraphics[width=0.999\textwidth]{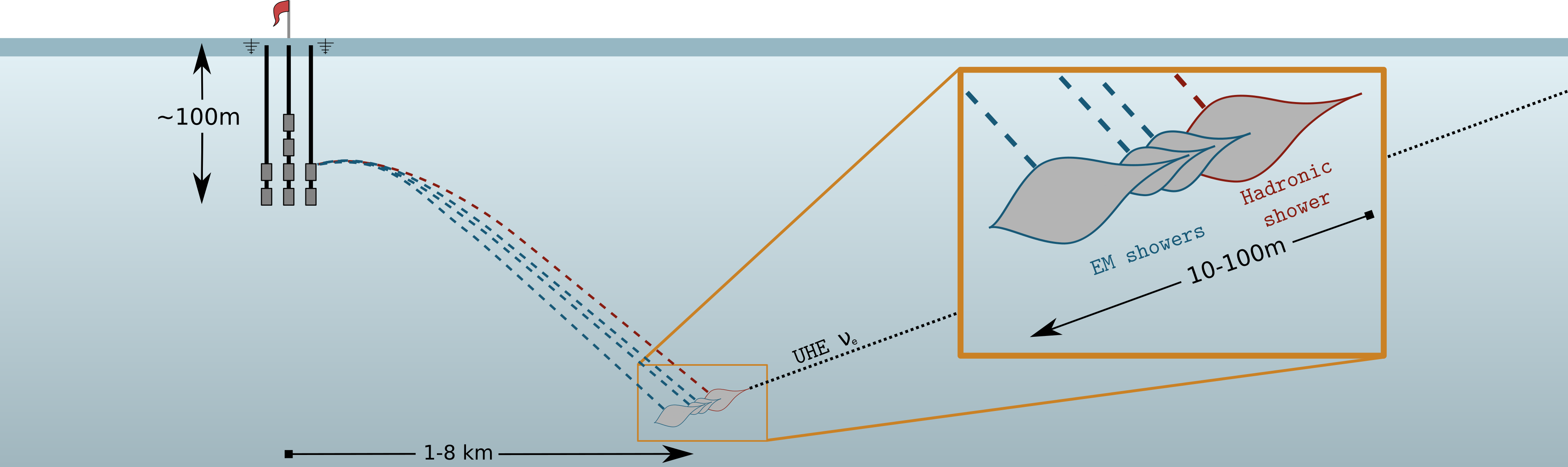}\\
 \vspace{0.2cm}
 \includegraphics[width=0.495\textwidth]{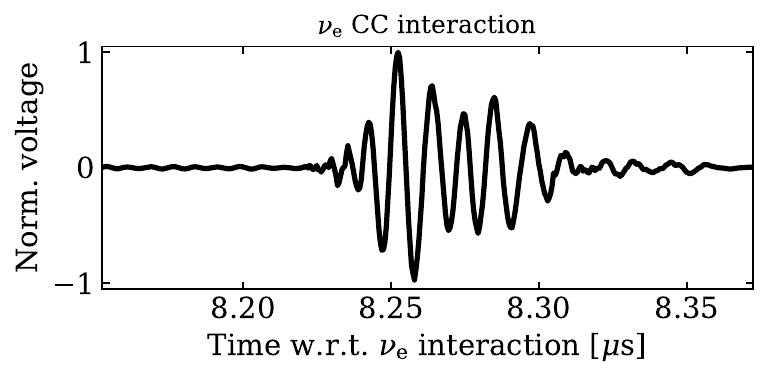}\hfill
 \includegraphics[width=0.495\textwidth]{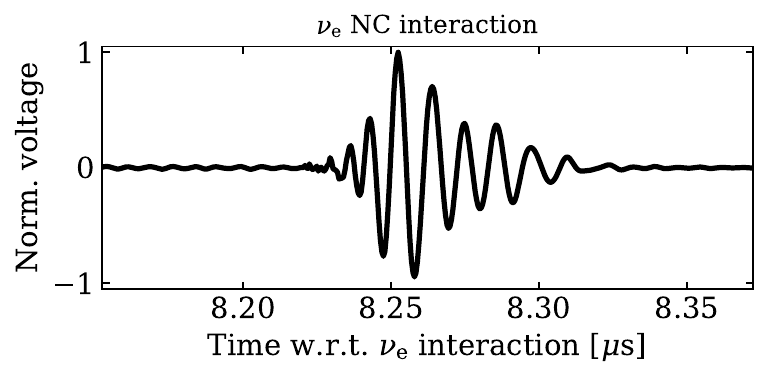}\\
 \vspace{0.1cm}
 \includegraphics[width=0.495\textwidth]{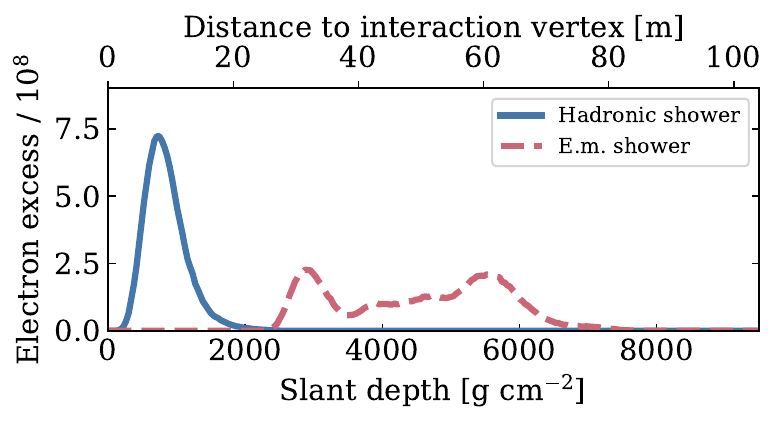}\hfill
 \includegraphics[width=0.495\textwidth]{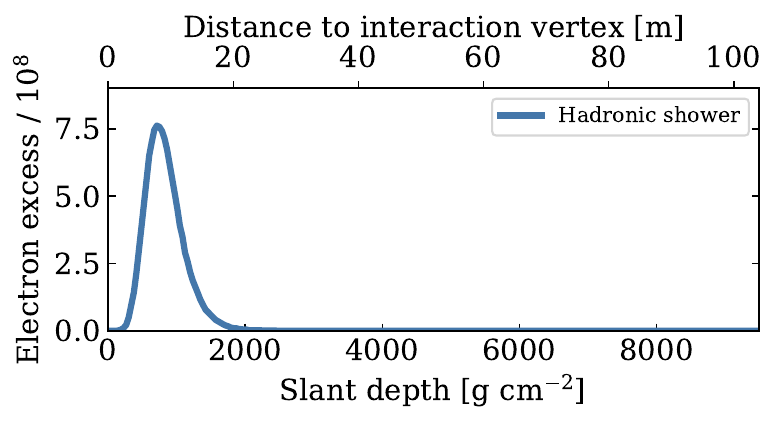}\\
 \caption{\textbf{\textit{Schematic of a charged-current interaction of a UHE \nue and its in-ice radio-detection.}}  {\it Top:} The primary neutrino-nucleon interaction produces a hadronic shower.  The final-state electron is affected by the LPM effect, resulting in multiple, separated electromagnetic (EM) sub-showers that imprint on the voltage waveform recorded by the detector stations.  The change of the index of refraction of radio with depth causes the trajectory of radio signals to bend on their way to the stations, an effect we account for in our simulations.  {\it Center:} Normalized noiseless voltage waveforms recorded in a detector antenna (an LPDA) triggered by a 3-EeV $\nu_e$ undergoing a CC ({\it left}) or NC ({\it right}) interaction, with otherwise identical initial conditions.  The additional late-time large-amplitude ``wiggles'' in the CC case, compared to the NC case, are due to the sub-showers induced by the LPM effect.  We build a custom convolutional neural network to identify their presence (\Cref{sec:cnn}).  {\it Bottom:}  Electron excess profiles as a function of the slant depth and geometric distance corresponding to the waveforms shown in the central panels, showing separately the contributions of the hadronic and electromagnetic showers.  See \Cref{sec:how_to_measure_flavor-lpm} for details.  \textit{Identifying $\nu_e$ CC events grants us access primarily to the UHE $\nu_e$ content; see \figu{observable_pairs}.}}
 \label{fig:lpm_waveform}
\end{figure*}

Figure~\ref{fig:lpm_waveform} shows how, due to the Landau-Pomeranchuk-Migdal (LPM) effect~\cite{Landau:1953um,Landau:1965ksp,Migdal:1956tc}, \nue CC interactions can be distinguished from all other neutrino interactions~\cite{Gerhardt:2010bj,Klein:2020nuk,Stjarnholm:2021xpj,ericsson2021investigations}. The LPM effect reduces the bremsstrahlung and pair-production cross sections of the high-energy electron generated in a \nue CC interaction, as a result of the interference between consecutive scatterings with the medium. This leads to delayed and stochastic shower development with potentially several slightly displaced sub-showers, which, together, elongates the particle profile of the shower and of its radio emission compared to the compact hadronic showers generated in other types of interactions~\cite{Glaser:2019cws,Barwick:2022vqt}. By identifying this elongation in detected showers, we gain sensitivity to the \nue content of the neutrino flux, as we explain below.

The magnitude of the LPM effect grows with energy. Below \SI{e18}{eV}, it mostly just delays the start of the electron-induced shower with respect to the hadronic shower. At higher energies, in addition, the electron-induced shower displays more structure and consists of multiple displaced sub-showers; see, \eg, \figu{lpm_waveform}, bottom left, and  \Refe~\cite{Barwick:2022vqt}. However, despite the shower elongation, the combined hadronic and electromagnetic showers remain small, \ie, a few tens of meters, and so they are detected in superposition by a single station.
The interference between two or more spatially separate, but overlapping or adjacent showers is encoded in the radio waveform recorded by the station~\cite{Alvarez-Muniz:2011wcg}. These subtle features in the waveform shape allow us to distinguish events that are initiated by \nue CC interactions from all other interactions, which in turn allows us to estimate the \nue fraction in the UHE neutrino flux. Also, pulse shapes generated by the \numu and \nutau CC interactions look like non-\nue CC interactions as they produce spatially displaced non-overlapping showers (see also the discussion of potential background below). 

In our analysis, we use a deep neural network to classify \nue CC {\it vs.}~non-\nue CC events; we describe it in \Cref{sec:cnn}. The classification is not perfect, \ie, showers due to \nue CC interactions can only be identified on a statistical basis and with limited accuracy, which limits the sensitivity of this detection channel. But, on the other hand, in a flux with equal content of all neutrino flavors, \nue CC interactions dominate the number of observed events; see \figu{event_rates_per_flavor} (also, Fig.~14 in \Refe~\cite{Valera:2022ylt}). Later, in \Cref{sec:results}, we report that this detection channel provides sensitivity to \nue similar to the sensitivity to $\numu + \nutau$ provided by the multi-shower channel. Yet, in contrast, the \nue CC detection channel has room for improvement. A better classifier would improve the sensitivity to \nue, whereas the multi-shower detection channel is limited by the underlying physics, \ie, by the probability of large stochastic energy losses of muons and taus. 

A background to the \nue CC detection channel would be any process that generates multiple overlapping showers in the ice whose radio emission interferes with each other, mimicking the interference due to the LPM effect. However, since our classifier is not perfect (\Cref{sec:cnn}), a small background contamination is acceptable and can be quantified and accounted for, statistically. Below, we briefly discuss potential backgrounds.

In the CC interaction of a \numu or \nutau, there is negligible probability that a large shower is initiated by the secondary interaction of a muon or tau close enough to the hadronic shower at the $\nu N$ interaction vertex for the radio emission from both interactions to interfere and mimic the waveform of a \nue CC interaction; cf.~\Refes~\cite{Garcia-Fernandez:2020dhb,Glaser:2021hfi}. Still, such secondary interaction---or decay, in the case of a tau (see below)---can initiate an EM shower. Yet, as this is only a single EM shower---as opposed to \nue CC interactions, where there is an additional hadronic shower with which to interfere---its energy would need to be fairly high for it to exhibit the sub-structure, \ie, multiple slightly-displaced sub-showers, that would set it apart from a hadronic shower and lead it to being incorrectly classified as coming from a \nue CC interaction. Thus, this represents a background only if the initial neutrino interaction was not also observed.
Hence, the potential background is the small fraction of high-energy EM showers induced by the already rare muons or taus whose secondary interactions are detected but whose primary interactions are missed.
Therefore, the background to \nue CC interactions from secondary muon and tau interactions is small; we will ignore it in the following.  

Regarding tau decays, they are only a relevant contribution to the detection rate up to approximately $5 \cdot 10^{17}$~eV, since higher-energy taus are too long-lived to decay inside the detector array, and only 18\% of those that do produce an electron~\cite{ParticleDataGroup:2022pth}. Therefore, the background to \nue CC interactions from tau decays is also small and we will ignore it. 

\begin{table*}[tb]
 \caption{\textbf{\textit{Topology of the neural network used to distinguish waveforms that are produced via showers influenced by the LPM effect.}} The size of the data in the (antenna, time, layer) dimensions, where applicable, are given along with the total number of trainable parameters for each block or layer.  See \Cref{sec:cnn-topology} for details.}
 \label{tab:network_topology}
 \renewcommand{\arraystretch}{1.1}
 \begin{ruledtabular}
  \begin{tabular}{lcccc}
   Network component & Size of the output data & Kernel size & Number of filters & Trainable parameters\\
   \hline
   Convolution block 1 & (5, 512, \phantom{1}32) & 5 & 32 & \phantom{1\,1}20\,800\\
   Convolution block 2 & (5, 128, \phantom{1}64) & 5 & 64 & \phantom{1\,1}71\,936\\
   Convolution block 3 & (5,\phantom{1} 32, 128) & 5 & 128 & \phantom{1}\,287\,232\\
   Convolution block 4 & (5,\phantom{11} 8, 256) & 5 & 256 & 1\,147\,904\\
   \hline
   Dense layer 1 & 512 & $\cdots$ & $\cdots$ & 1\,311\,232\\
   Dense layer 2 & 512 & $\cdots$ & $\cdots$ & \phantom{1}\,262\,656\\
   Dense layer 3 & 1 & $\cdots$ & $\cdots$ & \phantom{1\,11}1\,026\\
  \end{tabular}
 \end{ruledtabular}
\end{table*}


\subsection{Additional detection channels}
\label{sec:how_to_measure_flavor-additional_channels}

In the following, we briefly describe alternative detection channels that offer flavor sensitivity.  We do not explore them here, but they are explored elsewhere or might be explored in the future.

\begin{description}[style=unboxed]
 \item[Tau regeneration through the Earth]
  The Earth is opaque to UHE neutrinos, but UHE \nutau can propagate greater distances due to tau regeneration~\cite{Ritz:1987mh, Halzen:1998be, Alvarez-Muniz:2018owm,Garg:2022ugd,Cummings:2023dbj,Cummings:2023iuw}. The short lifetime of the tau causes it to decay back into a neutrino before losing a large fraction of its initial energy to the medium. This process can repeat several times, allowing \nutau to propagate farther through the Earth than \nue and \numu, but losing energy during this process. Therefore, a part of the \nutau flux at ultra-high energies is converted to a flux with energies two-to-three orders of magnitude lower, depending on the arrival direction to the detector~\cite{Safa:2019ege, Garcia:2020jwr,Safa:2021ghs,Valera:2022ylt}; most are expected to arrive from directions close to the horizon. Hence, the detection of Earth-traversing neutrinos provides a unique signature for \nutau. 
  However, the additional event rate from tau regeneration is small. Even for the huge in-ice radio detector foreseen for IceCube-Gen2, we expect less than one detectable event from tau regeneration in ten years of operation~\cite{Valera:2022ylt,Valera:private}. Therefore, we do not consider this detection channel in this work as it would not add additional sensitivity compared to the other two detection channels described above. A search for Earth-traversing neutrinos with optical neutrino telescopes might be more promising~\cite{IceCube:2021pue}.
 \item[Angular distribution of neutrinos]
  Reference~\cite{Wang:2013njo} proposed to infer the flavor composition of UHE neutrinos from the directional distribution of detected events. Different neutrino flavors have slightly different angular distributions, primarily because the probability of detecting a secondary interaction from muons and taus is highest for horizontal directions, which allows the lepton to propagate through more ice, thereby increasing the chance of undergoing a stochastic energy loss observed by a radio detector station. The work of reference~\cite{Wang:2013njo} from 10 years ago showed promising results, but is now dated, since it used overly optimistic neutrino flux models and fairly simplistic simulation codes compared to present-day standards, so that the results are not directly comparable to our work. It would be interesting to reassess this observable using, \eg, the state-of-the-art lepton propagation code {\tt PROPOSAL}~\cite{Koehne:2013gpa} and its integration into {\tt NuRadioMC}~\cite{Garcia-Fernandez:2020dhb}, on which our work is based. 
 \item[Combination of detection techniques]
  The all-flavor neutrino energy spectrum measured with in-ice radio detectors can be combined with the measurement of air-shower detectors, which are only sensitive to \nutau, to gain sensitivity to the \nutau content. However, the method requires the existence of two large UHE neutrino telescopes, which adds another layer of challenge. A potential future option would be the combination of measurements by IceCube-Gen2---all-flavor---and GRAND---only \nutau---as studied in \Refe~\cite{Testagrossa:2023ukh}.
\end{description}


\section{Identification of \mathintitle{\nu_e} CC interactions using a neural network}
\label{sec:cnn}

As described above, the LPM effect imprints characteristic features in neutrino-induced in-ice radio waveforms that allow us to identify UHE \nue CC events.  We developed a neural network that identifies these features, which we use to infer the fraction of \nue in the flux of UHE neutrinos.


\subsection{Simulated data}
\label{sec:cnn-sim_data}

\begin{figure*}
 \centering
 \includegraphics[width=\textwidth]{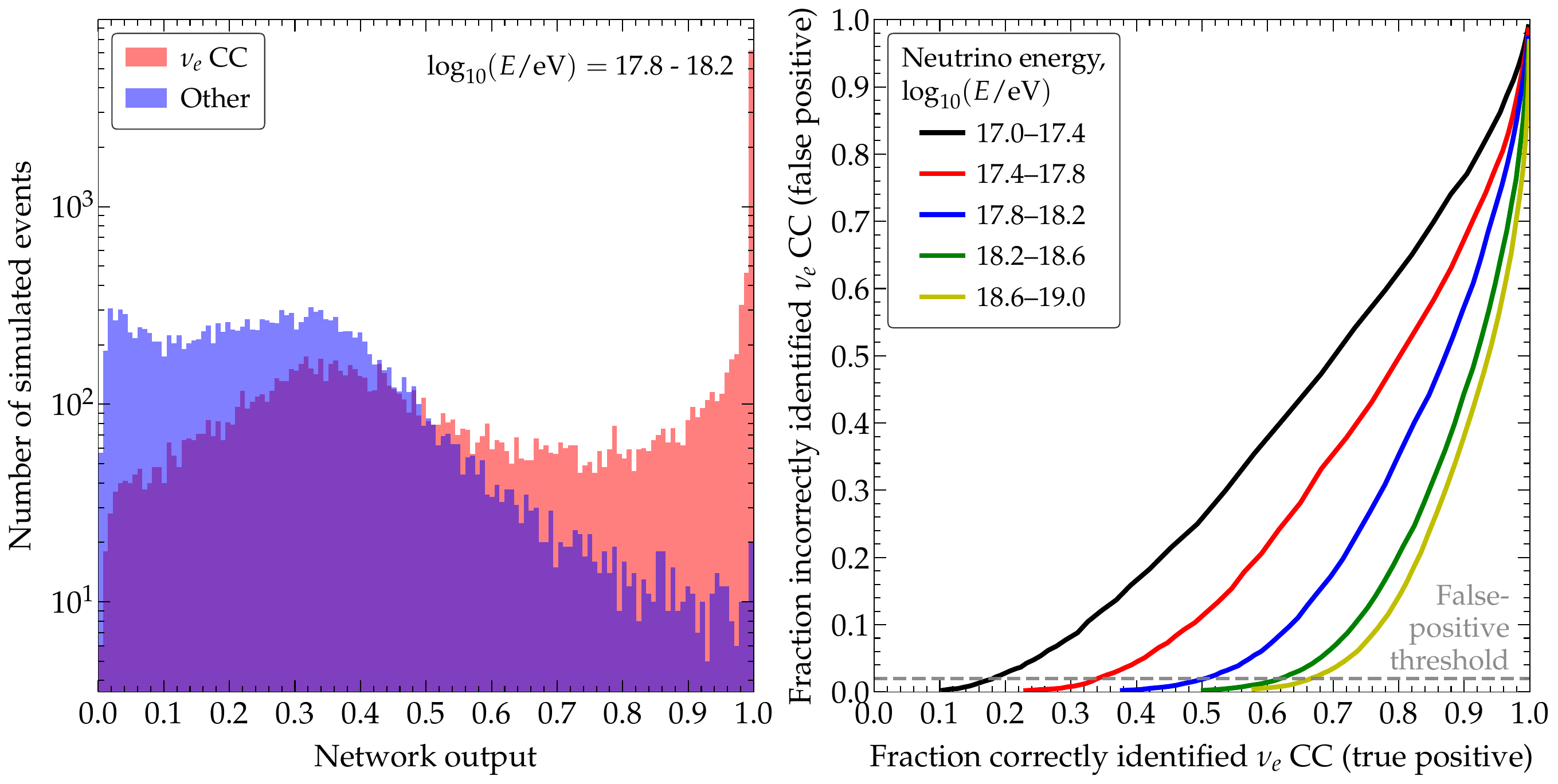}
 \caption{\textbf{\textit{Performance of the neural network in classifying \nue~CC events on a test data set.}} \textit{Left:} Network output of the final dense layer for \nue CC interactions and other interactions in the test data set, for one illustrative energy bin. \textit{Right:} True-positive and false-positive fractions for the identification of simulated \nue CC interactions, for different energy bins. The chosen false-positive rate for this analysis is 2\%. See \Cref{sec:cnn} for details. \emph{The classification is better at higher energies, where $>$\,65\% of \nue CC events are correctly identified.}}
 \label{fig:roc}
\end{figure*}

\begin{figure}[b!]
 \centering
 \includegraphics[width=\columnwidth]{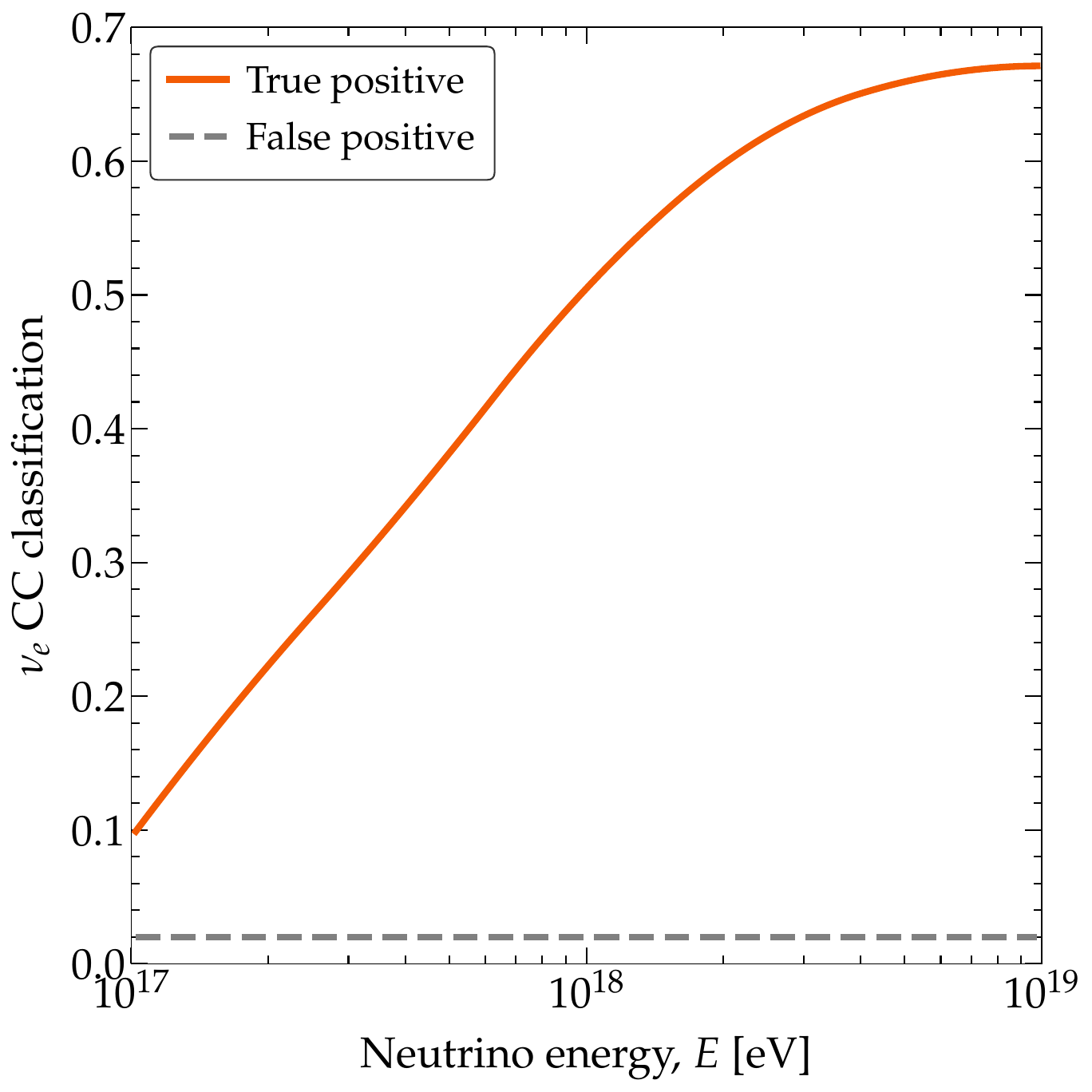}
 \caption{\textbf{\textit{True and false positive rates of our convolutional neural network to classify UHE \nue CC events.}}  We choose the threshold on the network output such that the false-positive rate is 2\%, which we found optimizes flavor identification. See \Cref{sec:cnn} for details.}
 \label{fig:fraction_nue_cc_classification}
\end{figure}

To train the network, we simulated the radio emission from in-ice cascades in the South Pole ice using {\tt NuRadioMC}~\cite{Glaser:2019rxw,Glaser:2019cws,Garcia-Fernandez:2020dhb,Glaser:2021hfi}.
Training was carried out for \emph{shallow stations} as pioneered by ARIANNA~\cite{Anker:2019rzo}, \ie, detector stations made up of four log-periodic dipole array (LPDA) antennas and one vertically polarized dipole antenna, buried a few meters in the ice. The configuration and trigger settings are the same as foreseen for IceCube-Gen2~\cite{IceCube-Gen2:2021rkf, IceCube-Gen2-TDR} and include a full trigger simulation in the presence of thermal noise. The simulations are identical to those in \Refe~\cite{Glaser:2022lky}; details can be found therein. 

The current baseline design for IceCube-Gen2 consists of a hybrid array of \emph{shallow} and \emph{deep} station components; a deep component comprises antennas buried up to a depth of \SI{150}{m} installed in three narrow boreholes~\cite{IceCube-Gen2-TDR}. In the following, we will assume that the results for shallow stations are representative for the entire IceCube-Gen2 array. The \nue CC sensitivity comes from the pulse shape, which we expect to be measured similarly well with deep detector components. Preliminary results of a deep neural network trained on simulated data of a deep station component indeed show similar performance and agree with the result of a shallow station that we presented in the following within a few percent.  As it may be more than ten years until the measurement we propose here can be conducted, and because in that time the final layout of IceCube-Gen2 is likely to evolve, we forego detailed studies of the impact of the array layout on the flavor sensitivity for now.

To train the network, we simulated  \nue with equal numbers of CC and NC interactions, for neutrino energies from $10^{17}$\,eV to $10^{19}$\,eV, which yield a representative sample of the two different event morphologies that the network aims to classify: \nue CC and non-\nue CC. 


\subsection{Neural network topology}
\label{sec:cnn-topology}

\Cref{tab:network_topology} summarizes the layout of the convolutional neural network (CNN) that we trained on the simulated data described in~\Cref{sec:cnn-sim_data} to distinguish \nue CC interactions from all other interactions.  The network takes as input the raw waveforms from the four LPDAs and the shallow vertical dipole antenna, sampled at 2~Gsamples~s$^{-1}$, resulting in 512 samples, each lasting 256~ns.
The encoding is performed using four blocks of convolutions. Each block consists of four convolutional layers followed by a max-pooling layer. Between each block, the max-pooling decreases the size of the time dimension by a factor of four while the number of convolutional filters increases by a factor of two such that the overall size of the data decreases after each block. 

\begin{figure*}[t!]
 \centering
 \includegraphics[width=\textwidth]{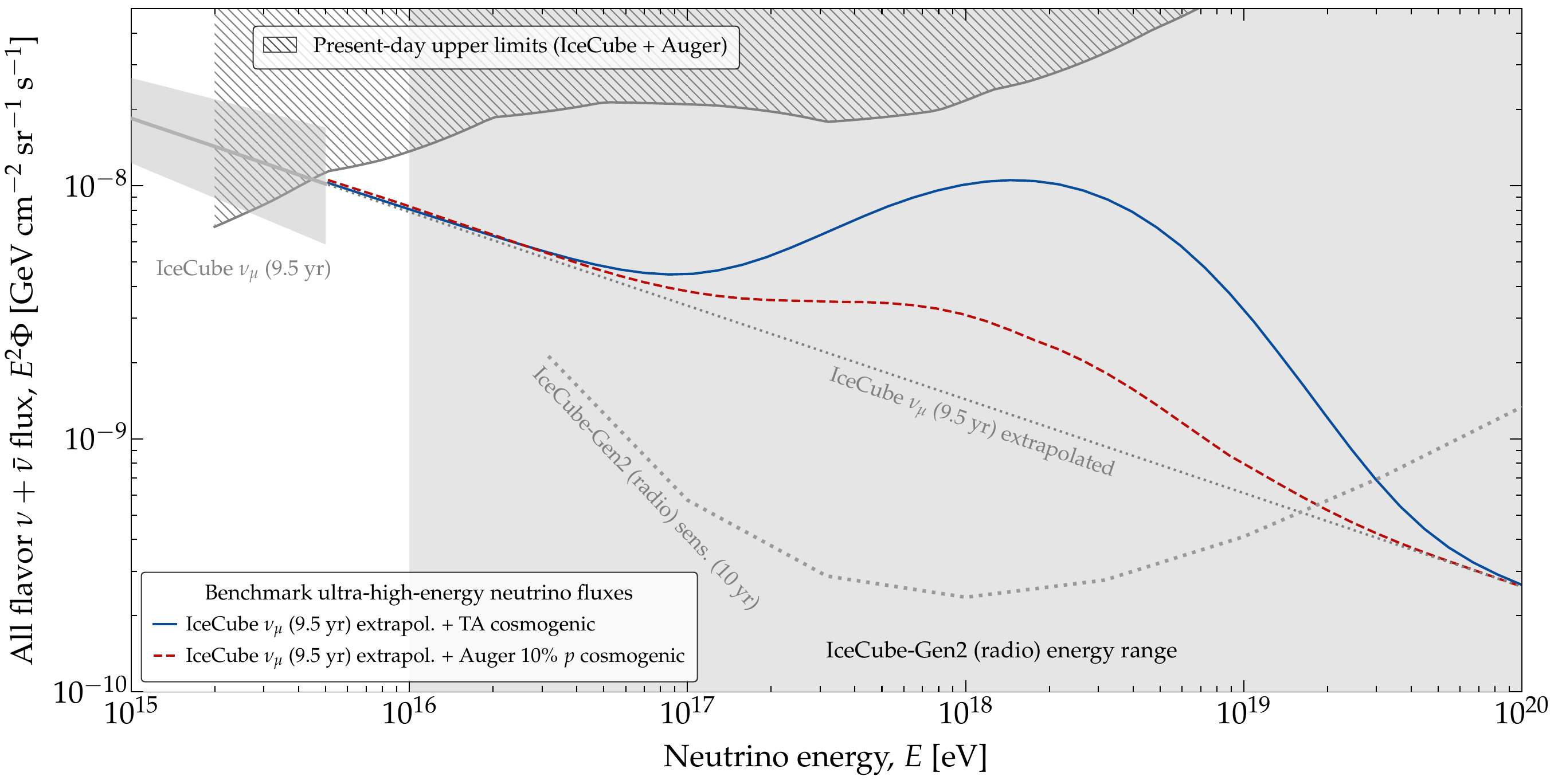}
 \caption{\textbf{\textit{Benchmark models of the UHE neutrino flux.}}  Our high-flux benchmark model is built from the cosmogenic flux derived from fits to UHECR data from the Telescope Array (TA)~\cite{Bergman:2021djm}; our low-flux one, from the cosmogenic flux derived from fits to UHECR data from the Pierre Auger Observatory (Auger), assuming that 10\% of the cosmic rays are protons~\cite{vanVliet:2019nse}.  To each, we add the IceCube flux derived from 9.5~years of through-going tracks~\cite{IceCube:2021uhz} (``IceCube \numu (9.5~yr)''), extrapolated to ultra-high energies.  The projected sensitivity of the radio array of IceCube-Gen2 is from \Refe~\cite{IceCube-Gen2:2021rkf}.  Present-day upper limits on the flux of UHE neutrinos are from IceCube~\cite{IceCube:2018fhm} and Auger~\cite{PierreAuger:2019ens}.  See \Cref{sec:sensitivity_flux} for details. \textit{Our benchmark flux models are representative of the breadth of theoretical flux predictions.}}
 \label{fig:fluxes}
\end{figure*}

During the convolution steps, a sliding kernel with five trainable weights is used to perform the convolution operation. Each antenna is treated independently---\ie, the convolution is one-dimensional, applied only to one antenna at a time---and the kernel weights are shared across the antennas. The output is then run through three fully connected---or \emph{dense}---layers. Ultimately, the CNN outputs a classification value for the event.  
This architecture was inspired by the Visual Geometry Group (VGG) architecture~\cite{Simonyan:2014cmh} and is motivated by the ability of convolution layers to efficiently analyze pulse shapes. A series of convolution layers with pooling in-between allows modeling more and more complex features. Variants of this architecture have been successfully used for the reconstruction of the neutrino direction and energy of UHE neutrinos \cite{Glaser:2022lky,IceCube-Gen2:2023wcw}, and for a low-level trigger \cite{Arianna:2021vcx,RNO-G:2023oxb}. 
Reference~\cite{ericsson2021investigations} contains further details about the CNN and how we arrived at this architecture.

\Cref{fig:roc} shows how the trained network performs on an independent test data set. We show the network output for \nue-CC and non-\nue-CC interactions with neutrino energies between $10^{17.8}$~eV and $10^{18.2}$~eV. The two distributions are distinct, but partially overlap.  By choosing a threshold value for the network output, we can optimize the true-positive rate, \ie, the fraction of correctly identified \nue CC events, against the false-positive rate, \ie, the fraction of \nue NC events that contaminate the data set.  We optimized this cut and found that flavor sensitivity is maximized for small false-positive values of around 0.5--4\%. The exact optimal value depends on the total number of events and the true flavor composition, and could be further fine-tuned to test specific hypotheses. In this generic analysis, we fix the false-positive fraction to 2\% for all hypotheses tested.

\Cref{fig:roc} right shows the resulting joint distributions of  true-positive and false-positive classified events, for all the energy bins used in our analysis (more on this in \Cref{sec:sensitivity_samples}).  The discrimination power of the network is indicated by how closely the distribution is to the bottom-right corner, \ie, to most \nue CC events being correctly classified with the least contamination from NC interactions.  The network performs better at higher energies, since the intensity of the LPM effect increases with energy.  Further, the average signal-to-noise ratio also increases with energy, meaning that the shape of the radio pulse is less affected by thermal fluctuations.  We impose a threshold on the network output such that the false-positive rate is 2\% in each of the energy bins. This sets the true-positive rate of the network to roughly 10\%--65\%, depending on the neutrino energy.

Figure~\ref{fig:fraction_nue_cc_classification} shows the resulting fractions of true-positive and false-positive classifications of our CNN as functions of the neutrino energy.  Because the ratio of CC to NC interactions is about 70\%, above $10^{18.5}$~eV half of all \nue interactions are correctly identified as such.

\begin{table*}
 \caption{\textbf{\textit{Benchmark scenarios of UHE neutrino flavor composition used in our forecasts.}}  For each neutrino flavor, $f_{\alpha, {\rm S}}$ is the ratio of $\nu_\alpha + \bar{\nu}_\alpha$ ($\alpha = e, \mu, \tau$) to the all-flavor neutrino flux emitted at the source.  After oscillations en route to Earth, the flavor composition is transformed into $f_{\alpha, \oplus}$.  We compute the latter assuming the present-day best-fit values of the neutrino mixing parameters from NuFit~5.2~\cite{Esteban:2020cvm, NuFit5.2}.  See \Cref{sec:sensitivity_flavor} for details.}
 \label{tab:flavor_ratios}
 \renewcommand{\arraystretch}{1.3}
 \begin{ruledtabular}
 \begin{tabular}{ccc}
  \multirow{2}{*}{Benchmark neutrino production} &
  \multicolumn{2}{c}{UHE neutrino flavor composition} \\
  &
  At sources, $(f_{e, {\rm S}}, f_{\mu, {\rm S}}, f_{\tau, {\rm S}})$ &
  At Earth, $(f_{e, \oplus}, f_{\mu, \oplus}, f_{\tau, \oplus})$ \\
  \hline
  Pion decay &
  $\left( \frac{1}{3}, \frac{2}{3}, 0 \right)$ &
  $\left( 0.30, 0.36, 0.34 \right)$ \\
  Muon-damped &
  $\left( 0, 1, 0 \right)$ &
  $\left( 0.17, 0.45, 0.37 \right)$ \\
  Neutron decay &
  $\left( 1, 0, 0 \right)$ &
  $\left( 0.55, 0.17, 0.28 \right)$ \\
 \end{tabular}
 \end{ruledtabular}
\end{table*}


\section{UHE neutrinos: predictions and detection}
\label{sec:sensitivity}

Using the techniques described above, we predict the sensitivity of IceCube-Gen2 to the flavor composition of UHE neutrinos.  We make projections for two plausible choices of the neutrino flux---a high one and a low one---and for three different benchmark choices of the flavor composition. The benchmark flux models are chosen because they have a shape that is representative of several predictions of UHE neutrino production, which have a generally similar shape. The benchmark flavor compositions are selected due to their astrophysical motivation and because they span the at-Earth phase space allowed by the standard model \footnote{While the flux models used in this study imply a specific flavor ratio at the source, they are decoupled in this work and are simply used as proxies to test how the sensitivity depends on the absolute flux normalization.}.  In the following, we describe these choices, the methods that we use to compute the rate of detected events in IceCube-Gen2, and our statistical methods.


\subsection{The all-flavor UHE neutrino flux}
\label{sec:sensitivity_flux}

The diffuse UHE neutrino flux is unmeasured, but there are multiple theoretical predictions of it; see, \eg, \Refes~\cite{Fang:2013vla, Padovani:2015mba, Fang:2017zjf, Heinze:2019jou, Muzio:2019leu, Rodrigues:2020pli, Muzio:2021zud} for examples and \Refe~\cite{Valera:2022ylt, Valera:2022wmu} for an overview.  The models are for cosmogenic neutrinos, produced in UHECR interactions on cosmological photon backgrounds during their propagation---\ie, the cosmic microwave background and the extragalactic background light---for source neutrinos, produced in UHECR interactions with matter and radiation in their sources, or for combinations of both.  

For our forecasts, it makes little difference what the origin of the UHE neutrino flux is; rather, what matters is the neutrino energy distribution.  Among the theory predictions, the neutrino energy spectrum varies significantly in size and shape, which is partially why planning for its discovery has been challenging; see, \eg, Fig.~2 in \Refe~\cite{Valera:2022wmu}.  Still, the predicted energy spectra share a few common features: most of them consist, roughly, of the superposition of a power-law in neutrino energy, especially towards lower energies, and a bump-like spectrum whose width and position vary depending on the model.  In some models, the power-law comes from neutrino production via proton-proton interactions, and the bump, from production via proton-photon interactions.  Below, we adopt two benchmark UHE neutrino flux models that exhibit these features, which are well motivated, and that are representative of the wide breadth of predictions.

\Cref{fig:fluxes} shows our two benchmark UHE neutrino flux models.  Each benchmark contains two flux components.

The first flux component is based on the neutrino flux inferred from TeV--PeV through-going muon tracks detected over 9.5~years by IceCube~\cite{IceCube:2021uhz}. In this energy range, the $\nu_\mu + \bar{\nu}_\mu$ spectrum is described by a power law $\propto E^{-2.37}$. We extrapolate it to ultra-high energies without changes to the shape of its energy spectrum, and multiply it by a factor of 3 to convert it to an all-flavor flux, under the nominal assumption that neutrinos of all flavors are equally abundant (\Cref{sec:sensitivity_flavor}).  Later, we re-distribute the all-flavor flux among different flavors using different assumptions of the flavor composition.  We include this flux component to account for the possibility that the TeV--PeV neutrino flux extends to higher energies.  This component yields 30--35 detected neutrinos over ten years, depending on its flavor composition.  

The second flux component is a prediction of the cosmogenic neutrino flux, inferred from fits to the measured UHECR spectrum and mass composition.  This is the dominant component at ultra-high energies. To build our benchmark fluxes, we consider two alternatives. Our high-flux benchmark uses the cosmogenic neutrino production derived from fits to UHECR data from the Telescope Array (TA)~\cite{Bergman:2021djm}; our low-flux one, the cosmogenic neutrino production derived from fits to UHECR data from the Pierre Auger Observatory (Auger), assuming that 10\% of the cosmic rays are protons~\cite{vanVliet:2019nse}.  While the original construction of these flux models included predictions of the energy-dependent flavor composition at Earth, for our analysis we use only the all-flavor flux.  Later, we re-distribute it among different flavors using different assumptions of the flavor composition. This flux component yields 139--164 detected neutrinos over ten years for the high benchmark and 23--28 neutrinos for the low one, depending on the flavor composition.


\subsection{The UHE neutrino flavor composition}
\label{sec:sensitivity_flavor}

\Cref{tab:flavor_ratios} summarizes the three benchmark choices of the UHE neutrino flavor composition that we adopt to make our forecasts of flavor measurement.  There are other possibilities for the flavor composition with which neutrinos can be produced, including changes with neutrino energy (see, \eg, \Refes~\cite{Mehta:2011qb, Bhattacharya:2023mmp, Liu:2023flr} and \Cref{sec:discussion}), but we take these three as benchmarks, constant across the energies that we consider. Below, we expand on them.

The interactions of UHECRs with matter and radiation produce high-energy pions that, upon decaying, produce high-energy neutrinos, \ie, $\pi^+ \to \mu^+ + \nu_\mu$, followed by $\mu^+ \to e^+ + \nu_e + \bar{\nu}_\mu$, and their charged-conjugated processes.  Thus, the full pion decay chain yields a flavor composition at the neutrino sources (S) of $(f_{e, {\rm S}}, f_{\mu, {\rm S}}, f_{\tau, {\rm S}}) = \left( \frac{1}{3}, \frac{2}{3}, 0 \right)$, where $f_{\alpha, {\rm S}}$ is the ratio of $\nu_\alpha + \bar{\nu}_\alpha$ to the all-flavor neutrino yield ($\alpha = e, \mu, \tau$).  We do not separate $\nu_\alpha$ from $\bar{\nu}_\alpha$ because high-energy neutrino telescopes presently cannot distinguish between them.  
In sources that harbor intense magnetic fields, the intermediate muons might cool via synchrotron radiation, so that the only high-energy neutrinos emitted are directly from the decay of pions; in this case, the flavor composition is  $(0, 1, 0)_{\rm S}$.  
Separately, the beta-decay of neutrons and neutron-rich isotopes produces a pure-$\bar{\nu}_e$ flux; in this case, the flavor composition is $(1, 0, 0)_{\rm S}$.  

En route to Earth ($\oplus$), neutrino oscillations modify the flavor composition into $f_{\alpha, \oplus} = \sum_{\beta=e,\mu,\tau} P_{\beta\alpha} f_{\beta, {\rm S}}$, where $P_{\beta\alpha} = \sum_{i=1}^3 \lvert U_{\beta i} \rvert^2 \lvert U_{\alpha i} \rvert^2$ is the average flavor-transition probability and $U$ is the Pontecorvo-Maki-Nakagawa-Sakata mixing matrix.  The latter depends on the neutrino mixing parameters $\theta_{12}$, $\theta_{23}$, $\theta_{13}$, and $\delta_{\rm CP}$.  \Cref{tab:flavor_ratios} shows the flavor composition at Earth for our three benchmark cases, computed using the present-day best-fit values of the mixing parameters.  Production by pion decay yields close to flavor equipartition at Earth; this is the nominal expectation.

Today, the uncertainty in the mixing parameters is sizable and impacts the prediction of the allowed neutrino flavor composition at Earth~\cite{Bustamante:2015waa, Song:2020nfh, Liu:2023flr}.  However, in the years 2030--2040, the mixing angles should be known to within 1\% of their values and the CP violation phase should be known to within a few percent of its value, thanks to measurements by upcoming oscillation experiments DUNE, Hyper-Kamiokande, and JUNO.  At that point, the impact of the uncertainties in the mixing parameters on the predicted flavor composition at Earth should be tiny~\cite{Song:2020nfh}.  Therefore, in our forecasts below, we neglect the uncertainty in the mixing parameters, and fix them to their present-day best-fit values from the NuFit~5.2 global fit to oscillation data~\cite{Esteban:2020cvm, NuFit5.2}.


\subsection{Generating samples of detected events}
\label{sec:sensitivity_samples}

\begin{figure}[t!]
 \centering
 \includegraphics[width=\columnwidth]{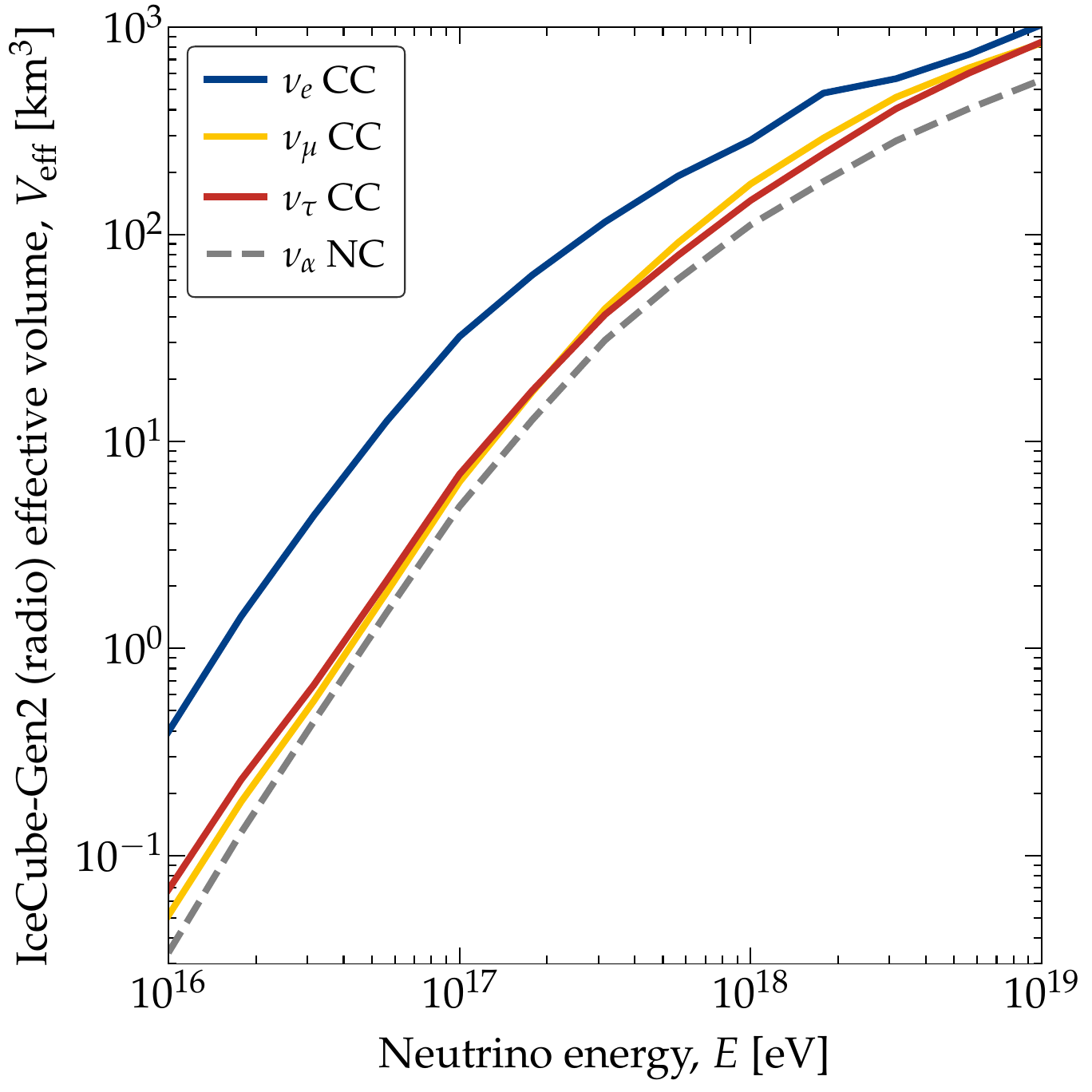}
 \caption{\textbf{\textit{Effective volume for UHE neutrino detection in the radio array of IceCube-Gen2.}}  We treat separately CC interactions of \nue, \numu, and \nutau, and NC interactions of all neutrino flavors.  See \Cref{sec:sensitivity_samples} for details.}
 \label{fig:effective_volume}
\end{figure}

To generate measurement forecasts of the UHE neutrino flavor composition, we generate mock samples of events detected by a radio array based on the IceCube-Gen2 design~\cite{IceCube-Gen2-TDR}.  To do this, we estimate the response of the detector via its simulated effective volume.

\Cref{fig:effective_volume} shows the effective volumes that we use. They are calculated using {\tt NuRadioMC}~\cite{Glaser:2019cws,Glaser:2019rxw}, using the same settings as for other IceCube-Gen2 studies; see \Refes~\cite{IceCube-Gen2:2021rkf,IceCube-Gen2-TDR} for details. The effective volumes are direction-averaged and include the effect of the attenuation of the neutrino flux while propagating inside the Earth, i.e., the probability of a neutrino reaching the simulation volume. 
We approximate the effective volume of the full detector by multiplying the effective volume of a single shallow detector station and a single deep detector station by the respective number of shallow and deep stations foreseen in the IceCube-Gen2 radio array, as was done in previous studies that forecast its science potential; \eg, \Refes~\cite{Valera:2022wmu,Valera:2022ylt,Valera:2023ayh,Fiorillo:2022ijt}).  
Doing this slightly overestimates the effective volume at the highest energies, since in reality a fraction of $\nu N$ interactions is expected to be detected by multiple stations. This fraction depends on the station spacing. The design goal of IceCube-Gen2 is to keep the fraction at about 10\% at \SI{e18}{eV}~\cite{IceCube-Gen2-TDR}. We ignore this correction in the following. 

We calculate the effective volume separately for \nue CC and NC interactions.  Then we use the results of \Refe~\cite{Glaser:2021hfi} to increase the \numu and \nutau CC effective volumes by accounting for the observation of showers produced by the muons and taus. The NC effective volumes of all flavors are the same.  Because in \nue CC interactions the entire neutrino energy is converted into electromagnetic energy, the effective volume for \nue CC interactions dominates over all other channels; \eg, at \SI{e17}{eV}, it is larger than the other channels by a factor of about 5.  

\begin{figure*}[t!]
 \centering
 \includegraphics[width=\textwidth]{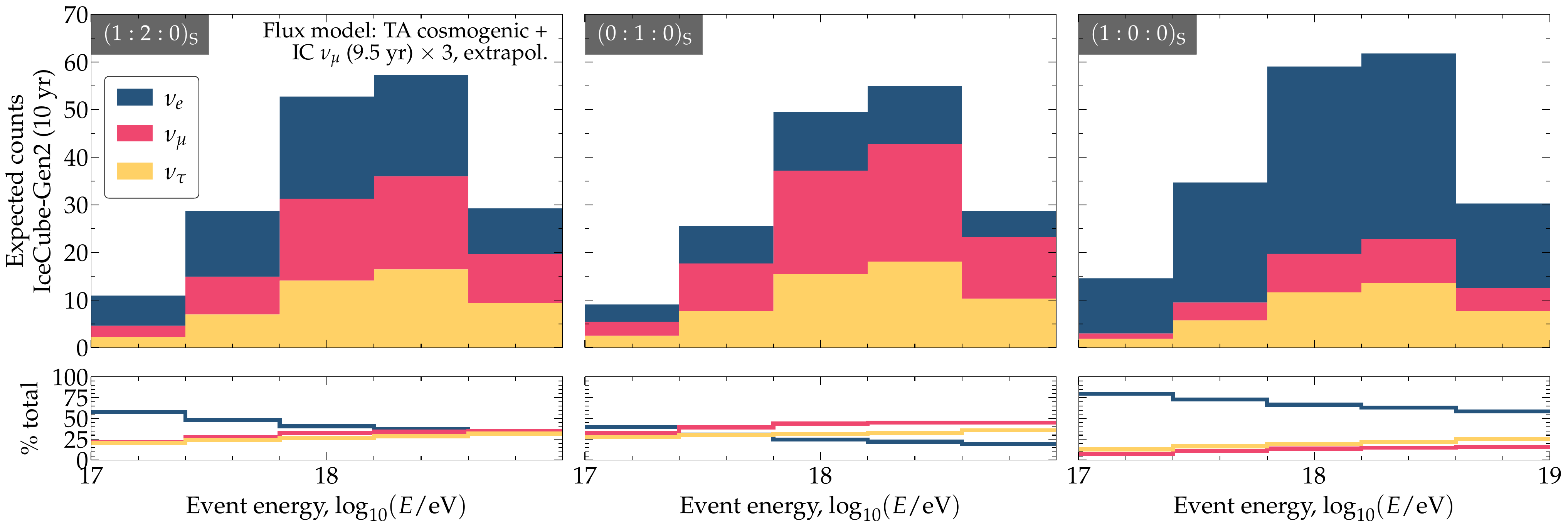}
 \caption{\textbf{\textit{Mean expected number of detected events in the radio array of IceCube-Gen2, showing contributions of each neutrino flavor.}}  For this figure (also \figu{event_rates_per_method}), we assume our high UHE neutrino benchmark flux model, which uses the cosmogenic neutrino prediction based on TA UHECR data (\figu{fluxes}). To produce our forecasts, we also compute the spectrum assuming our lower UHE neutrino benchmark flux model (not shown here).  {\it Left:} Assuming the production of UHE neutrinos from the decay of pions at the sources, $(1/3, 2/3, 0)_{\rm S}$, which yields $ (0.30, 0.36, 0.34)_\oplus$ at Earth. {\it Center:}  Assuming muon-damped flavor composition at the sources, $(0, 1, 0)_{\rm S}$, which yields $(0.17, 0.45, 0.37)_\oplus$.  {\it Right:} Assuming neutrino production via neutron decay at the sources, $(1, 0, 0)_{\rm S}$, which yields $ (0.55, 0.17, 0.28)_\oplus$.  See \Cref{sec:sensitivity_samples} for details.}
 \label{fig:event_rates_per_flavor}
\end{figure*}

Using the above effective volumes, $V_\alpha^{\rm CC}$ and $V_\alpha^{\rm NC}$, we compute the effective areas, $A_\alpha^{\rm CC} \equiv V_\alpha^{\rm CC} / \lambda_\alpha^{\rm CC}$ and $A_\alpha^{\rm NC} \equiv V_\alpha^{\rm NC} / \lambda_\alpha^{\rm NC}$, where $\lambda_\alpha^{\rm CC} \equiv (\sigma_{\nu N}^{\rm CC} n_N )^{-1}$ is the neutrino CC interaction length in ice, $\sigma_{\nu N}^{\rm CC}$ is the $\nu N$ CC cross section and $n_N$ is the number density of nucleons in ice, and similarly for NC.  At these energies, the CC cross section of $\nu_\alpha$ and $\bar{\nu}_\alpha$ of all flavors are nearly equal, and the same holds for NC. Thus, for a given choice of all-flavor UHE neutrino flux, $\Phi_{\nu_{\rm all}}$, and of the flavor composition at Earth, $f_{\alpha, \oplus}$, from among our benchmarks (\Cref{sec:sensitivity_flux} and \ref{sec:sensitivity_flavor}), the mean differential number of detected events due to $\nu_\alpha + \bar{\nu}_\alpha$ CC interactions, after an observation time $T$, is
\begin{equation}
 \frac{dN_{\nu_\alpha}^{\rm CC}}{dE}
 = 
 4 \pi
 T
 f_{\alpha, \oplus}
 \Phi_{\nu_{\rm all}}(E) A_\alpha^{\rm CC}(E)
 \;,
 \label{equ:events_by_type}
\end{equation}
and similarly for NC, with CC~$\to$~NC.  The factor of $4\pi$ in \equ{events_by_type} accounts for integration over the full sky.  We integrate \equ{events_by_type}, and its NC counterpart, in energy to compute the event rates in $N_E = 5$ energy bins between $10^{17}$\,eV and $10^{19}$\,eV, evenly spaced in logarithmic scale, \ie, $N_{\nu_\alpha, i}^{\rm CC}$ and $N_{\nu_\alpha, i}^{\rm NC}$ in the $i$-th bin.  
The energy range in this analysis is selected by the overlap of where the CNN was trained and where the data to define the effective area was available. This energy range also contains the bulk of the expected events for the benchmark flux models, which drive the sensitivity to the flavor.
Our main forecasts are for $T = 10$~years of detector exposure. Later, as part of our statistical analysis (\Cref{sec:results_nuecc,sec:results_multi,sec:results_total}), we account for statistical fluctuations in the number of detected events by using the mean event rate as the central value of a Poisson distribution.

\Cref{fig:event_rates_per_flavor} shows the contribution of the different neutrino flavors to the mean energy spectrum of events, for our three benchmark choices of flavor composition.  The event rate is highest between $10^{18}$~eV and $10^{18.6}$~eV because this is where our benchmark UHE neutrino flux peaks (\figu{fluxes}).  At lower energies, the effective volume is small; at higher energies, the flux is low.  Under approximate flavor equipartition at Earth, \ie, the pion-decay model with $\left( 1/3, 2/3, 0 \right)_\mathrm{S}$ and $\left( 0.30, 0.36, 0.34 \right)_\oplus$, the dominant contribution is from \nue, since its effective volume is the largest (\figu{effective_volume}).  For other choices of the flavor composition, the contribution of the different flavors reflects the underlying flavor composition. 

\begin{figure*}[t!]
 \centering
 \includegraphics[width=\textwidth]{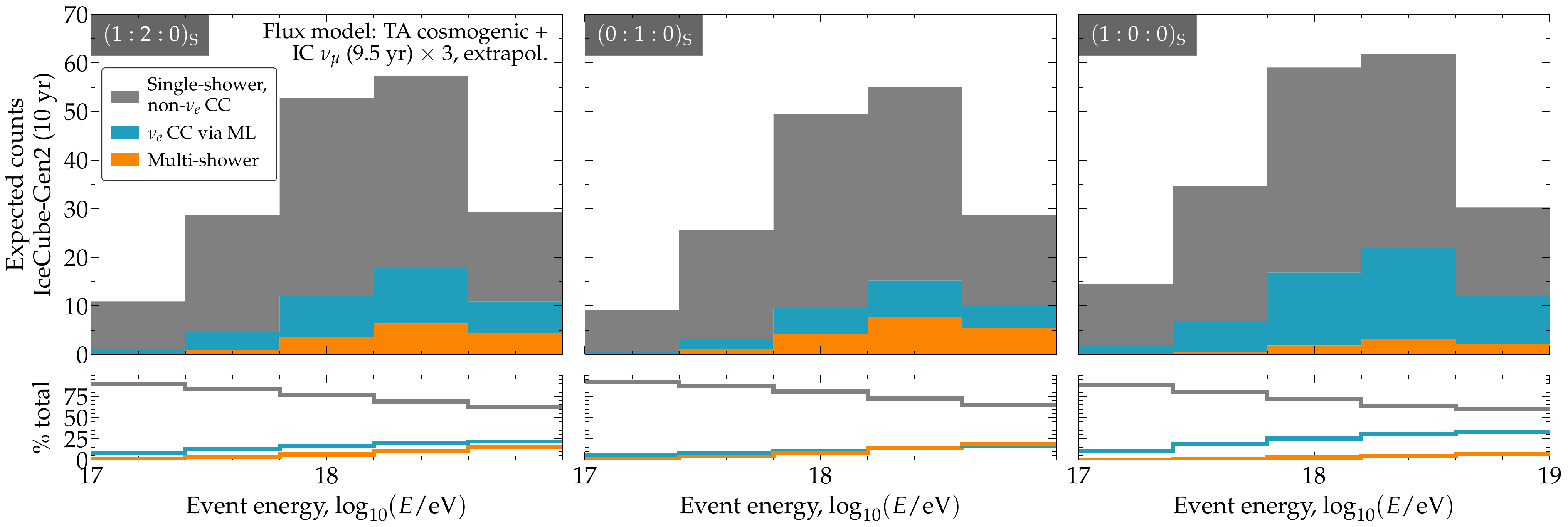}
 \caption{\textbf{\textit{Mean expected numbers of observed events, showing contributions of each detection channel.}}  Same as \figu{event_rates_per_flavor}, but showing instead the contributions of each detection channel: single-station events not classified as $\nu_e$ CC, single-station events classified as $\nu_e$ CC (\Cref{sec:how_to_measure_flavor-lpm}), and multi-shower events (\Cref{sec:how_to_measure_flavor-multi_shower}). See \Cref{sec:sensitivity_samples} for details.}
 \label{fig:event_rates_per_method}
\end{figure*}

We compute three event energy spectra, or observables, that we use in our analysis below:
\begin{enumerate}
 \item
  {\bf Total event spectrum, $N_{\nu_{\rm all}, i}$:}
  This is the number of events due to the CC plus NC interactions of \nue, \numu, and \nutau, \ie, $N_{\nu_{\rm all}, i} = N_{\nu_e, i}^{\rm CC} + N_{\nu_\mu, i}^{\rm CC} + N_{\nu_\tau, i}^{\rm CC} + ({\rm CC} \to {\rm NC})$.  It does not separate flavors.
 \item 
  {\bf Spectrum of identified \nue CC events, $N_{{\rm ML}, i}$:}
  We compute the number of \nue CC events identified by the CNN, $N_{{\rm ML}, i}$ (``ML'' refers to ``machine learning''), by adding together a randomly chosen subset of the number of \nue CC events, $N_{\nu_e, i}^{\rm CC}$, selected according to the true-positive fraction of the CNN (see~\Cref{fig:roc}), and a randomly chosen subset of the number of events that are not \nue CC events, $N_{\nu_{\rm all}, i} - N_{\nu_e, i}^{\rm CC}$, selected according to the false-positive fraction of the CNN.
 \item
  {\bf Spectrum of multi-shower events, $N_{{\rm mult}, i}$:}
  We compute the number of events where two or more showers trigger multiple detector stations, $N_{\rm mult}$, by adding together a randomly chosen subset of the number of events triggered by \numu CC interactions, $N_{\nu_\mu, i}^{\rm CC}$, and a randomly chosen subset of the number of events triggered by \nutau CC interactions, $N_{\nu_\tau, i}^{\rm CC}$, based on the expected fraction of multi-shower events (\figu{fraction_multi_station}).
\end{enumerate}

\Cref{fig:event_rates_per_method} shows the above three event spectra.  Like for \figu{event_rates_per_flavor}, the overall shape of the event spectra  reflects the interplay of the neutrino energy spectrum and the detector effective volume.  Our capability to measure the UHE neutrino flavor composition rests on identifying \nue CC events and multi-shower events.  \Cref{fig:event_rates_per_method} reveals that these events make up less than 25\% of the total between $10^{18}$~eV and $10^{18.6}$~eV, where most of the events lie given our two flux benchmarks.  Below, we show that although these events are sub-dominant, they are common enough to grant us the flavor sensitivity that we seek. 


\section{Measurement forecasts of the UHE neutrino flavor composition}
\label{sec:results}


\subsection{Overview of the statistical analysis}
\label{sec:results_stat}

\begin{table*}[t!]
 \caption{\textbf{\textit{Free model parameters varied in our statistical analysis.}}  Each parameter floats independently of each other as part of our procedure to infer the UHE flavor composition. See \Cref{sec:results_stat} for details. }
 \label{tab:fitting_parameters}
 \begin{ruledtabular}
 \begin{tabular}{c|cc}
  \multicolumn{2}{c}{Model parameter} &
  Symbol      \\
  \hline
  \multicolumn{3}{c}{Physical parameters} \\
  \hline
  \multirow{2}{*}{UHE neutrino flavor composition} 
  &
  Electron flavor ratio &
  $f_{e, \oplus}$       \\
  &
  Muon flavor ratio     &
  $f_{\mu, \oplus}$     \\
  \hline
  \multicolumn{3}{c}{Nuisance parameters (profiled over)} \\
  \hline
  \multirow{4}{*}{\makecell{All-flavor UHE neutrino flux,\\
  $\log_{10}[\Phi_{\nu_{\rm all}}/({\rm GeV}^{-1}~{\rm cm}^{-2}~{\rm s}^{-1}~{\rm sr}^{-1})]$\\(third-degree polynomial in $x \equiv \log_{10}(E/{\rm eV})$)}} 
  &
  Coefficient $\propto x^3$ &
  $a$                \\
  &
  Coefficient $\propto x^2$ &
  $b$                \\
  &
  Coefficient $\propto x^1$ &
  $c$                 \\
  &
  Coefficient $\propto x^0$ &
  $d$                 \\
 \end{tabular}
 \end{ruledtabular}
\end{table*}

To gauge the flavor sensitivity of the in-ice radio array, first, we generate a mock sample of events assuming one of our benchmark choices of the all-flavor flux (\Cref{sec:sensitivity_flux}) and of the flavor composition (\Cref{sec:sensitivity_flavor}).  We take this as our observed event sample, consisting of $N_{\nu_{\rm all}, i}^{\rm obs}$, $N_{{\rm ML}, i}^{\rm obs}$, and $N_{{\rm mult}, i}^{\rm obs}$ (\Cref{sec:sensitivity_samples}).  Second, we perform a hypothesis test to recover from this sample the most likely flavor composition, \ie, $f_{e, \oplus}$, $f_{\mu, \oplus}$, and $f_{\tau, \oplus}$.  We do this by comparing, via a likelihood function (defined later), our observed event sample against test event samples computed under varying hypotheses.

Table~\ref{tab:fitting_parameters} shows the free model parameters of our analysis; they describe the all-flavor neutrino energy spectrum and the flavor composition.  For the all-flavor energy spectrum, we proceed as in an analysis of real data in which the normalization and shape of the all-flavor energy spectrum, $\Phi_{\nu_{\rm all}}$, would be unknown. Thus, we approximate the all-flavor spectrum by a third-order log-log polynomial in neutrino energy, which is able to capture the bump-like shape of our benchmark energy spectra (\figu{fluxes}).  It is described by the coefficients $a$, $b$, $c$, and $d$, whose values we let float in a comparison to the observed sample.  (Other parametrizations are possible, though we do not explore them here; see \Refe~\cite{Mena:2014sja}.) For the flavor composition, we let $f_{e, \oplus}$ and $f_{\mu, \oplus}$ float; since $f_{\tau, \oplus} \equiv 1 - f_{e, \oplus} - f_{\mu, \oplus}$, we do not float it separately.

In our analysis, we vary the values of $a$, $b$, $c$, $d$, $f_{e, \oplus}$, and $f_{\mu, \oplus}$ and, for each realization of them, we generate a test event sample consisting of $N_{\nu_{\rm all}, i}$, $N_{{\rm ML}, i}$, and $N_{{\rm mult}, i}$ (\Cref{sec:sensitivity_samples}). We compare the test event sample against our assumed observed sample via a likelihood function (defined below) that accounts for Poisson fluctuations in the number of events.  The likelihood is then marginalized with respect to the all-flavor spectrum parameters, and we report measurements on the flavor composition alone.

Below, we apply the procedure outlined above to three sets of observables to achieve sensitivity only to the \nue content via $N_{\nu_{\rm all}, i}$ plus $N_{{\rm ML}, i}$ (\Cref{sec:results_nuecc}), sensitivity only to the $\numu + \nutau$ content via $N_{\nu_{\rm all}, i}$ plus $N_{{\rm mult}, i}$ (\Cref{sec:results_multi}), and sensitivity to the three flavors via $N_{\nu_{\rm all}, i}$ plus $N_{{\rm ML}, i}$ plus $N_{{\rm mult}, i}$ (\Cref{sec:results_total}). 


\subsection{The \mathintitle{\nue} CC channel}
\label{sec:results_nuecc}

For a given test realization of the model parameters, $\boldsymbol{\theta} \equiv (a, b, c, d, f_{e, \oplus}, f_{\mu, \oplus})$ (Table~\ref{tab:fitting_parameters}), the likelihood function that compares the observed {\it vs.}~test events is 
\begin{equation}
 \label{equ:llh_ml}
 \mathcal{L}_{\rm ML}(\boldsymbol{\theta})
 =
 \prod_i^{N_E} 
 \mathcal{P}\left(\obs{N}_{\nu_{\rm all},i}; 
 N_{\nu_{\rm all},i}(\boldsymbol{\theta})\right)
 P_{{\rm ML}, i}(\boldsymbol{\theta}) \;,
\end{equation}
where a Poisson distribution, $\mathcal{P}$, accounts for the probability of observing $\obs{N}_{\nu_{\rm all},i}$ events when the expected number is $N_{\nu_{\rm all},i}(\boldsymbol{\theta})$.  The term $P_{{\rm ML}, i}$ accounts for the different ways in which the expected number of true-positive and false-positive events can sum up to $\obs{N}_{\rm{ML},i}$, \ie,
\begin{eqnarray}
 \label{equ:p_ml}
 P_{{\rm ML}, i}(\boldsymbol{\theta})
 &=&
 \sum_{k=0}^{\obs{N}_{\rm{ML},i}} 
 \mathcal{B}\left(
 k; \obs{N}_{\nu_{\rm all},i}, p_{{\rm tp},i}(\boldsymbol{\theta})
 \right)\nonumber\\
&\times &
 \mathcal{B}\left(
 \obs{N}_{\rm{ML},i}-k; \obs{N}_{\nu_{\rm all},i}, p_{{\rm fp},i}(\boldsymbol{\theta})
 \right) \;,
\end{eqnarray}
where $\mathcal{B}(k; N, p)$ is the binomial probability of observing $k$ out of $N$ events, each having a probability $p$ of being observed.  For true-positive (tp) and false-positive (fp) events, the latter is
\begin{eqnarray}
 \label{equ:p_tp}
 p_{{\rm tp},i}(\boldsymbol{\theta}) 
 &=&
 \frac{N_{\nu_e,i}^{\rm CC}(\boldsymbol{\theta})}
 {N_{\nu_{\rm all},i}(\boldsymbol{\theta})}
 \mathcal{T}_i \;, \\
 \label{equ:p_fp}
 p_{{\rm fp},i}(\boldsymbol{\theta})
 &=&
 \frac{N_{\nu_{\rm all},i}(\boldsymbol{\theta}) - N_{\nu_e,i}^{\rm CC}(\boldsymbol{\theta})}
 {N_{\nu_{\rm all},i}(\boldsymbol{\theta})}
 \mathcal{F}_i \;,
\end{eqnarray}
where $\mathcal{T}_i$ and $\mathcal{F}_i$ are, respectively, the true-positive and false-positive fractions of events classified by our CNN as being due to a \nue CC interaction (\figu{fraction_nue_cc_classification}), averaged in energy inside the $i$-th bin. 

For given values of $f_{e, \oplus}$ and $f_{\mu, \oplus}$, we profile the full likelihood, \equ{llh_ml}, by finding the values of the spectrum-shape parameters, $a$, $b$, $c$, $d$, that maximize it, \ie,
\begin{equation}
 \label{equ:llh_ml_red}
 \mathcal{L}_{\rm ML}(f_{e, \oplus}, f_{\mu, \oplus})
 =
 \max_{a,b,c,d}
 \mathcal{L}_{\rm ML}(\boldsymbol{\theta}) \;.
\end{equation}
This is the reduced likelihood from which we measure the flavor composition in this channel.  We report the best-fit flavor composition and the measurement uncertainties as confidence intervals of $f_{\alpha, \oplus}$, computed using Wilks' theorem.  We follow analogous procedures to make forecasts for other detection channels later.

\begin{figure}[t!]
 \centering
 \includegraphics[width=\columnwidth]{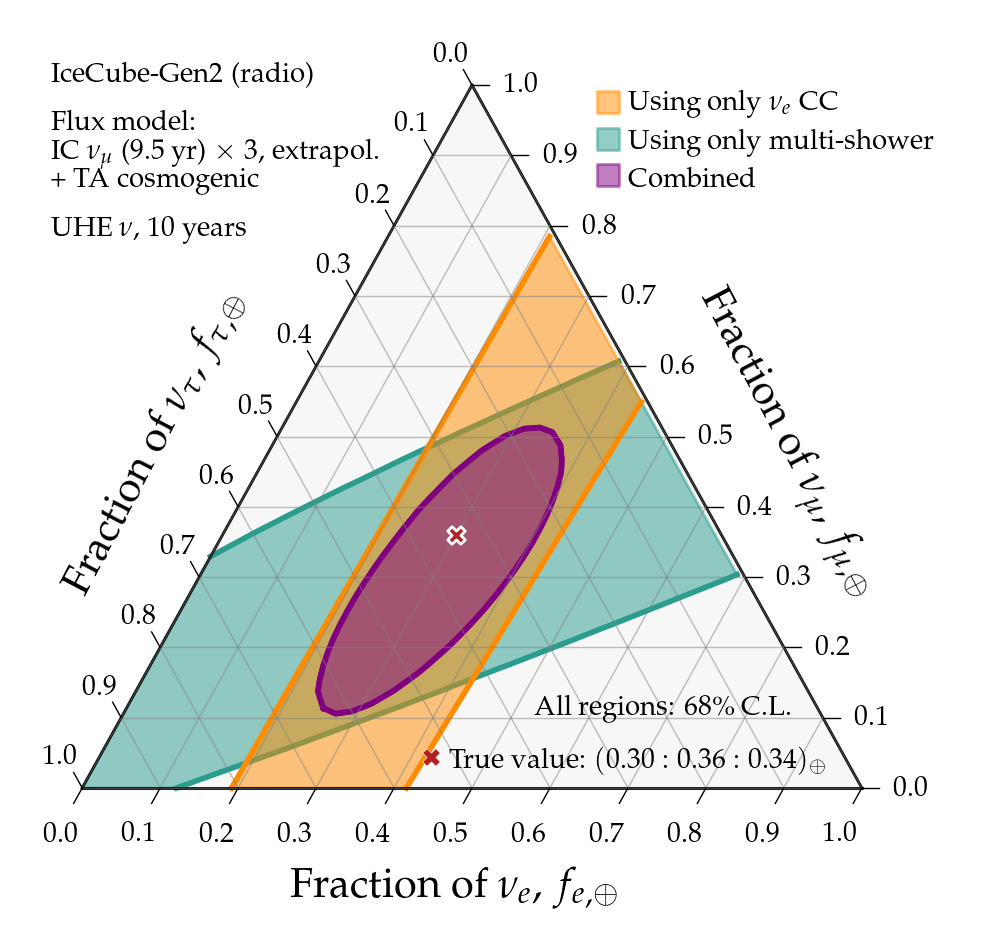}\\
 \includegraphics[width=\columnwidth]{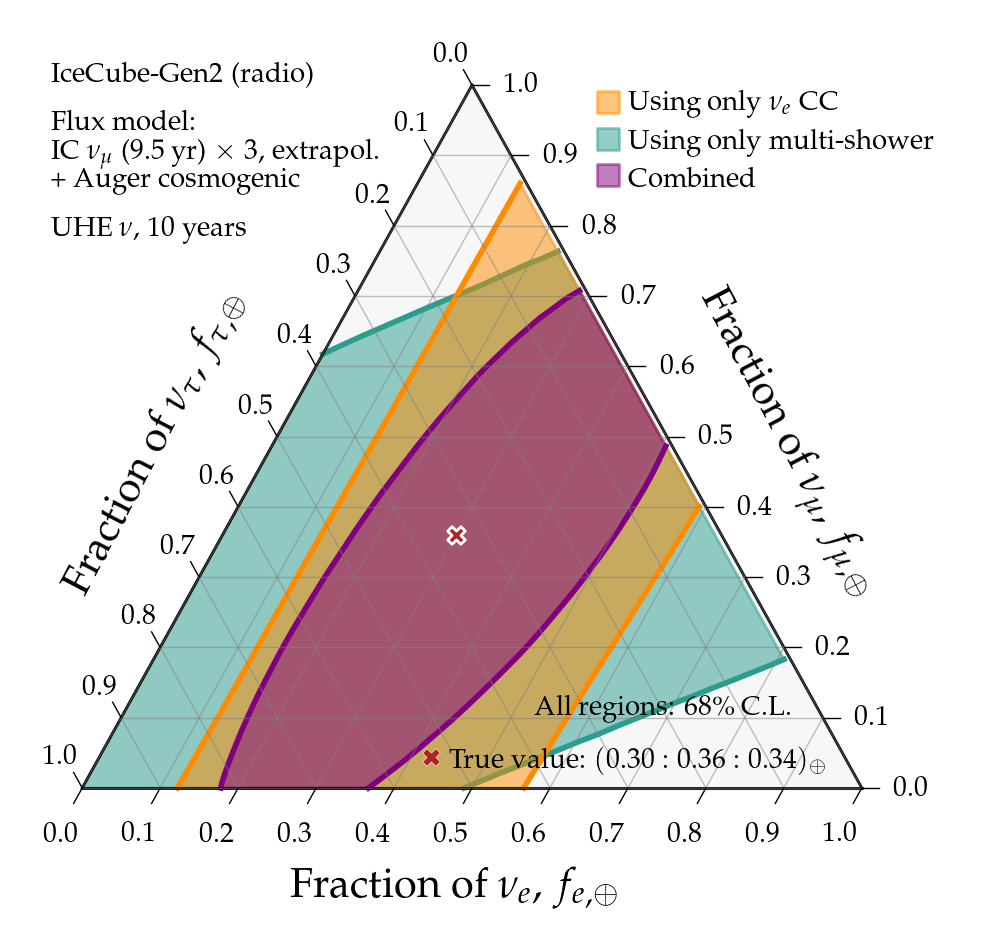}
 \caption{\textbf{\textit{Measured flavor composition of the diffuse flux of UHE neutrinos in the radio array of IceCube-Gen2, showing the contributions of the separate flavor-sensitive detection channels.}}  The regions represent the measurement uncertainty (68\%~C.L.) assuming neutrino production via pion decay.  {\it Top:} Results assuming our high all-flavor benchmark UHE neutrino flux, obtained from a fit to TA UHECR observations (\figu{fluxes}).  {\it Bottom:} Results assuming our low benchmark flux, obtained from a fit to Auger UHECR observations.  See Sections~\ref{sec:how_to_measure_flavor-lpm} and \ref{sec:how_to_measure_flavor-multi_shower} for a description of the two detection channels, \figu{ternary_contours_benchmarks} for results assuming other benchmarks of flavor composition, and \Cref{sec:results} for details.  \textit{The $\nu_e$~CC channel provides sensitivity to the $\nu_e$ content; the multi-shower channel provides sensitivity mainly to $\nu_\mu + \nu_\tau$ and also to the $\nu_e$ content.} 
 \label{fig:observable_pairs}}
\end{figure}

\begin{figure}[t!]
 \centering
 \includegraphics[width=\columnwidth]{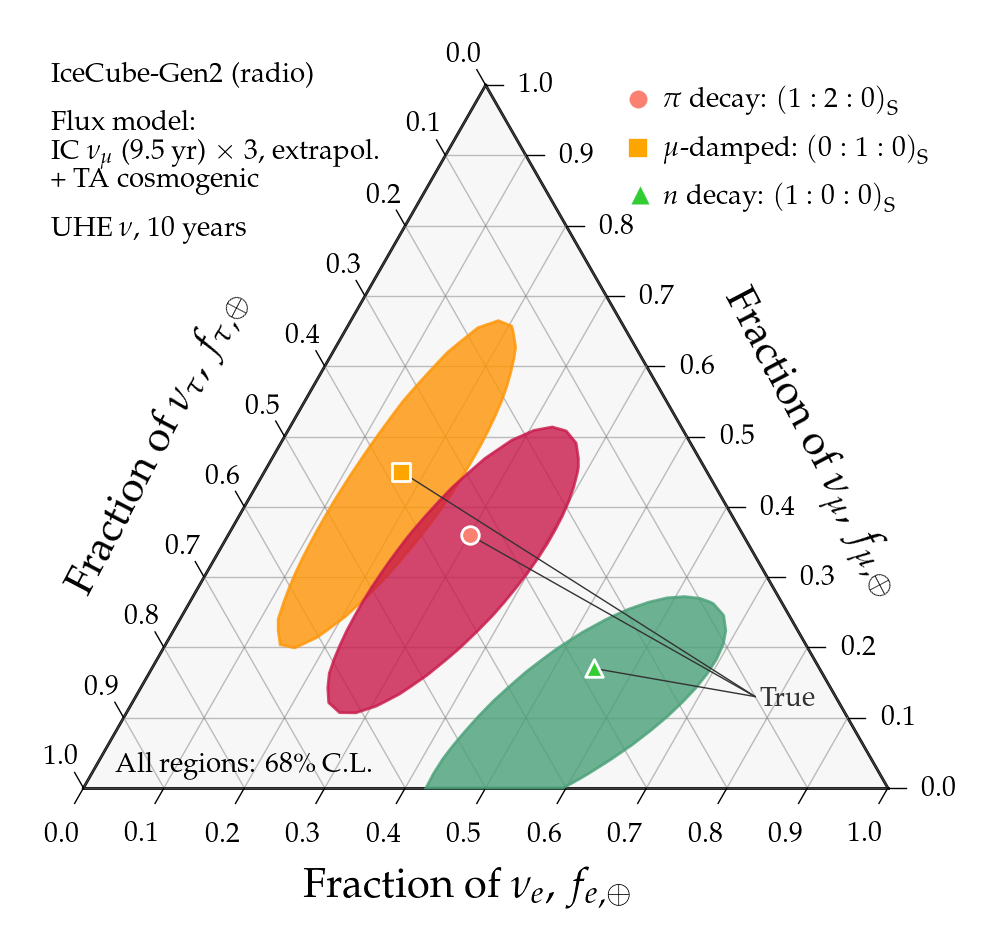}
 \includegraphics[width=\columnwidth]{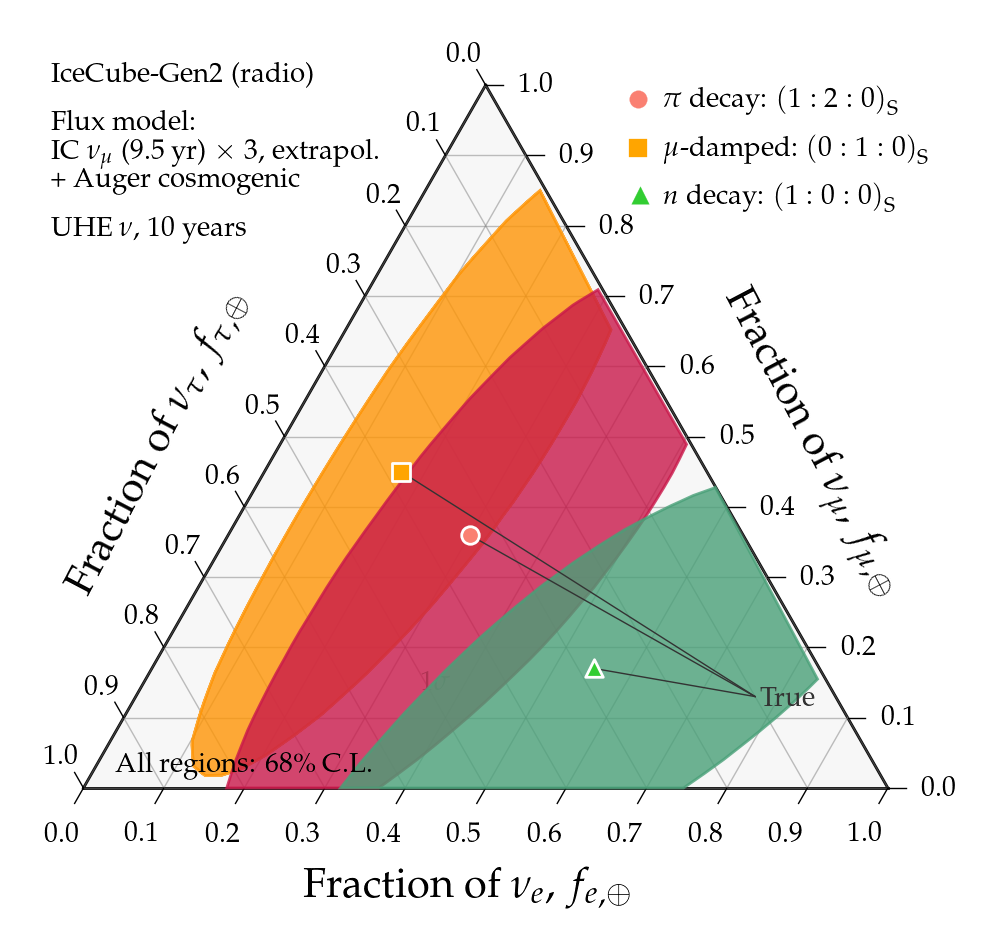}
 \caption{\textbf{\textit{Measured flavor composition of the diffuse flux of UHE neutrinos in the radio array of IceCube-Gen2.}}  The regions represent the measurement uncertainty (68\%~C.L.) assuming as true each of our three benchmark cases of flavor composition in turn (\Cref{sec:sensitivity_flavor}).  {\it Top:} Results assuming our high all-flavor benchmark UHE neutrino flux, obtained from a fit to TA UHECR observations (\figu{fluxes}).  {\it Bottom:} Results assuming our low benchmark flux, obtained from a fit to Auger UHECR observations.  See \figu{triangle_full} in Appendix~\ref{sec:appendix} for the full flavor likelihood and \Cref{sec:results} for details.  \textit{The in-ice radio-detection of UHE neutrinos may allow us, via our proposed methods (\Cref{sec:how_to_measure_flavor}), to measure their flavor composition, contingent on the size of their flux.}}
 \label{fig:ternary_contours_benchmarks}
\end{figure}

Figure~\ref{fig:observable_pairs} shows the resulting measurement of the flavor composition using $\nu_e$ CC events alone, assuming, for illustration, that its true value is that from the full pion decay chain.  Naturally, using \nue CC events alone allows us to measure only $f_{e, \oplus}$.  After 10 years of exposure, $f_{e, \oplus}$ could be measured to within 8\%, assuming our high-flux UHE neutrino benchmark model.  Fortunately, the theoretically palatable region of predicted flavor composition at Earth~\cite{Bustamante:2015waa, Song:2020nfh} (\figu{ternary_he_vs_uhe}) is aligned nearly orthogonally to the $f_{e, \oplus}$ axis, so that measuring $f_{e, \oplus}$ affords the greatest sensitivity to distinguish between predictions of the flavor composition at Earth, assuming standard oscillations.  

\begin{figure}[t!]
 \centering
 \includegraphics[width=\columnwidth]{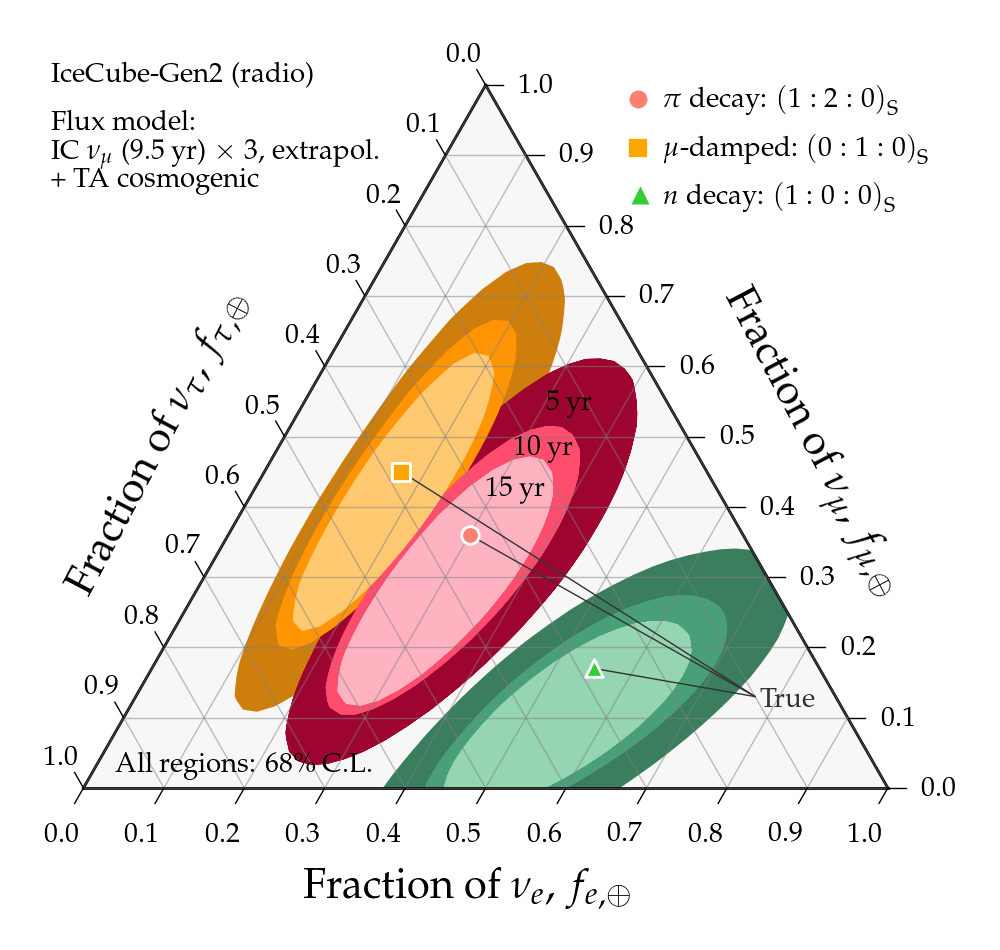}
 \caption{\textbf{\textit{Measured flavor composition of the diffuse flux of UHE neutrinos in the radio array of IceCube-Gen2,  varying detector exposure.}}  Similar to \figu{ternary_contours_benchmarks}, which was generated assuming 10 years of detector exposure, but comparing results of assuming 5, 10, and 15 years of exposure.  For this figure, we assume our high UHE neutrino benchmark flux.  See \figu{posterior_src_compare} for associated results on the inferred flavor composition at the sources. \textit{The increase in exposure from 10 to 15 years improves the precision with which flavor composition can be measured only marginally.}}
 \label{fig:ternary_contours_benchmarks_ta_compare}
\end{figure}


\subsection{The multi-shower channel}
\label{sec:results_multi}

Showers triggered by \numu and \nutau CC interactions can be identified due to muons and taus triggering multiple stations due to their stochastic energy losses (\Cref{sec:how_to_measure_flavor-multi_shower}). 

In this channel, the observables are the number of NC + CC events of all flavors, $\obs{N}_{\nu_{\rm all},i}$, and the number of multi-shower events, $\obs{N}_{{\rm mult},i}$.  Together, they provide sensitivity to the three flavors, primarily to \nue and to the combination of $\numu + \nutau$.  The sensitivity to \nue comes from the \nue CC effective volume being larger than for any other detection channel (\figu{effective_volume}). The sensitivity to $\numu + \nutau$ comes from detecting multi-shower events; the degeneracy between \numu and \nutau is only weakly mitigated due to the larger effective volume for \numu than \nutau for the multi-shower channel by about a factor of two (\figu{fraction_multi_station}).  By combining $\obs{N}_{\nu_{\rm all},i}$ with $\obs{N}_{{\rm mult},i}$, the degeneracy between a higher value of $f_{e,\oplus}$ and a lower all-flavor neutrino flux is lifted by the need to also correctly predict the number of multi-shower events.

In analogy with \Cref{sec:results_nuecc}, the likelihood function is
\begin{equation}
 \label{equ:llh_mult}
 \mathcal{L}_{\rm mult}(\boldsymbol{\theta})
 =
 \prod_i^{N_E}
 \mathcal{P}\left( \obs{N}_{\nu_{\rm all},i}; N_{\nu_{\rm all},i}(\boldsymbol{\theta}) \right) 
 P_{{\rm mult}, i}(\boldsymbol{\theta}) \;,
\end{equation}
where $\mathcal{P}$ is, like before, a Poisson distribution.  The probability $P_{{\rm mult}, i}$ accounts for the different ways in which the contributions of \numu and \nutau to the rate of multi-shower events can sum up to $N_{{\rm mult}, i}^{\rm obs}$, \ie,
\begin{eqnarray}
 \label{equ:p_multi}
 P_{{\rm mult}, i}(\boldsymbol{\theta})
 &=&
 \sum_{k=0}^{\obs{N}_{\rm mult}} 
 \mathcal{B} \left(
 k; 
 \obs{N}_{\nu_{\rm all},i}, 
 p_{\mu,i}(\boldsymbol{\theta})
 \right)\nonumber\\
 &\times&
 \mathcal{B}\left(
 \obs{N}_{\rm mult}-k; 
 \obs{N}_{\nu_{\rm all},i}, 
 p_{\tau,i}(\boldsymbol{\theta})
 \right) \;,
\end{eqnarray}
where $\mathcal{B}$ is, like before, the binomial distribution, $p_{\mu,i}$ and $p_{\tau,i}$ are, respectively, the probability of a \numu and \nutau CC interaction triggering a multi-shower event, \ie,
\begin{eqnarray}
 p_{\mu,i}(\boldsymbol{\theta}) 
 &=&
 \frac{N_{\nu_\mu,i}^{\rm CC}(\boldsymbol{\theta})}
 {N_{\nu_{\rm all},i}(\boldsymbol{\theta})}\,r_{\mu,i} \;, \\
 p_{\tau,i}(\boldsymbol{\theta})
 &=&
 \frac{N_{\nu_\tau,i}^{\rm CC}(\boldsymbol{\theta})}
 {N_{\nu_{\rm all},i}(\boldsymbol{\theta})}\,r_{\tau,i} \;,
\end{eqnarray}
and $r_{\mu,i}$ and $r_{\tau,i}$ are, respectively, the average fraction of \numu and \nutau CC interactions that trigger multi-shower events (\figu{fraction_multi_station}), averaged in energy inside the $i$-th bin. After profiling over the spectrum-shape parameters, the likelihood becomes $\mathcal{L}_{\rm mult}(f_{e, \oplus}, f_{\mu, \oplus})$.

Figure~\ref{fig:observable_pairs} shows the resulting measurement of the flavor composition using multi-shower events alone.  The measurement precision is comparable, but worse, than when using $\nu_e$ CC events alone.  The measurement region is approximately aligned along the line $f_{\mu, \oplus} + f_{\tau, \oplus} = 2/3$, the expectation from the pion-decay benchmark.  The offset from that is due to the fact that the multi-shower channel is also partially sensitive to the $\nu_e$ content.  Due to the tilt of this region, the multi-shower detection channel is less effective, by itself, in distinguishing between our three benchmarks of flavor composition.

The fraction of multi-shower events (\figu{fraction_multi_station}) depends on the station spacing---which we took to be 2~km---and on the station design, \eg, on using \emph{deep} versus \emph{shallow} detector components, or a combination of them. While for the bare discovery of UHE neutrinos~\cite{Valera:2022wmu} isolating a large fraction of multi-shower events could be considered a needless reduction of the effective volume of the detector, the potential to constrain the flavor composition depends on such design considerations.


\subsection{Total sensitivity}
\label{sec:results_total}

Finally, we combine all the available observables---the NC + CC events of all flavors, the events classified as \nue CC, and the multi-shower events---to find the full flavor sensitivity of the radio array of IceCube-Gen2.  Building on sections~\ref{sec:results_nuecc} and \ref{sec:results_multi}, the likelihood function is
\begin{eqnarray}
 \label{equ:llh_ml_mult}
 \mathcal{L}_{\rm ML+mult}(\boldsymbol{\theta})
 &=&
 \prod_i^{N_E}
 \mathcal{P}\left( \obs{N}_{\nu_{\rm all},i}; N_{\nu_{\rm all},i}(\boldsymbol{\theta}) \right)\nonumber\\
 &\times&
 P_{{\rm ML}, i}(\boldsymbol{\theta})
 P_{{\rm mult}, i}(\boldsymbol{\theta}) \;,
\end{eqnarray}
where $\mathcal{P}$, $P_{{\rm ML}, i}$, and $P_{{\rm mult}, i}$ are defined as before.  After profiling over the spectrum-shape parameters, this becomes $\mathcal{L}_{\rm ML+mult}(f_{e, \oplus}, f_{\mu, \oplus})$.

\Cref{fig:ternary_contours_benchmarks} shows our results, assuming our high and low benchmark fluxes and, separately, taking each of our three benchmarks choices of flavor composition as the true one.  As we pointed out in \Cref{sec:results_nuecc}, the strong constraining power in the $f_{e, \oplus}$ direction is advantageous, since that is the dimension along which our benchmarks are the most separated. This reaffirms that, contingent on the neutrino flux being high enough, $f_{e, \oplus}$ could be measured with precision enough to distinguish between our three benchmark choices of flavor composition at  68\%~C.L.~at the minimum---\eg, between full pion decay and muon-damped pion decay---to significantly more than 95\%~C.L.---\eg, between muon-damped pion decay and neutron decay.  

The addition of multi-shower events grants us sensitivity to the non-\nue content.  In \figu{ternary_contours_benchmarks}, it is what allows us to disfavor low and high content of \numu and \nutau, and to close the 68\%~C.L.~measurement contours, rather than being able only to measure bands of constant $f_{e, \oplus}$ or of approximately constant $f_{\mu, \oplus} + f_{\tau, \oplus}$, as in \figu{observable_pairs}.  While this does not significantly improve our capability to distinguish between our three flavor-composition benchmarks, it still reduces the measurement errors of the non-\nue content.  In addition, it allows us to probe predictions of the flavor composition with a low or high \numu or \nutau content, as posited by numerous proposed models of new neutrino physics (\Cref{sec:discussion}).

When repeating the analysis using our low benchmark flux model, \figu{ternary_contours_benchmarks} shows that the measurement uncertainty deteriorates due to the approximate factor-of-three reduction in the number of detected events. The larger uncertainties render the distinction between alternative choices of flavor composition unfeasible, except for the ones most different from each other, like muon-damped and neutron decay.  Unsurprisingly, like at lower energies, the measurement of the UHE flavor composition hinges on the flux being sufficiently large.  Still, \figu{ternary_contours_benchmarks} reveals that while a lower neutrino flux would practically remove any sensitivity to $f_{\mu, \oplus}$ and $f_{\tau, \oplus}$, it would only erode, but not destroy, the sensitivity to $f_{e, \oplus}$.

Figure~\ref{fig:ternary_contours_benchmarks_ta_compare} shows that increasing the detector exposure from 10 to 15 years only improves the precision of flavor measurements marginally.  We show how the results change under our low-flux benchmark model, when increasing the exposure from 10 to 20 years in \figu{ternary_contours_benchmarks_auger_compare}.  Ultimately, improving the precision beyond what is shown in  \figu{ternary_contours_benchmarks_ta_compare}, within a reasonable time frame, would require improving the identification of $\nu_e$ CC events or exploiting additional flavor-sensitive channels  or the combined detection by multiple UHE neutrino telescopes (\Cref{sec:how_to_measure_flavor-additional_channels}); see, \eg, Fig.~1 in \Refe~\cite{Testagrossa:2023ukh}.


\section{Applications}
\label{sec:discussion}

\begin{figure}[t]
 \centering
 \includegraphics[width=\columnwidth]{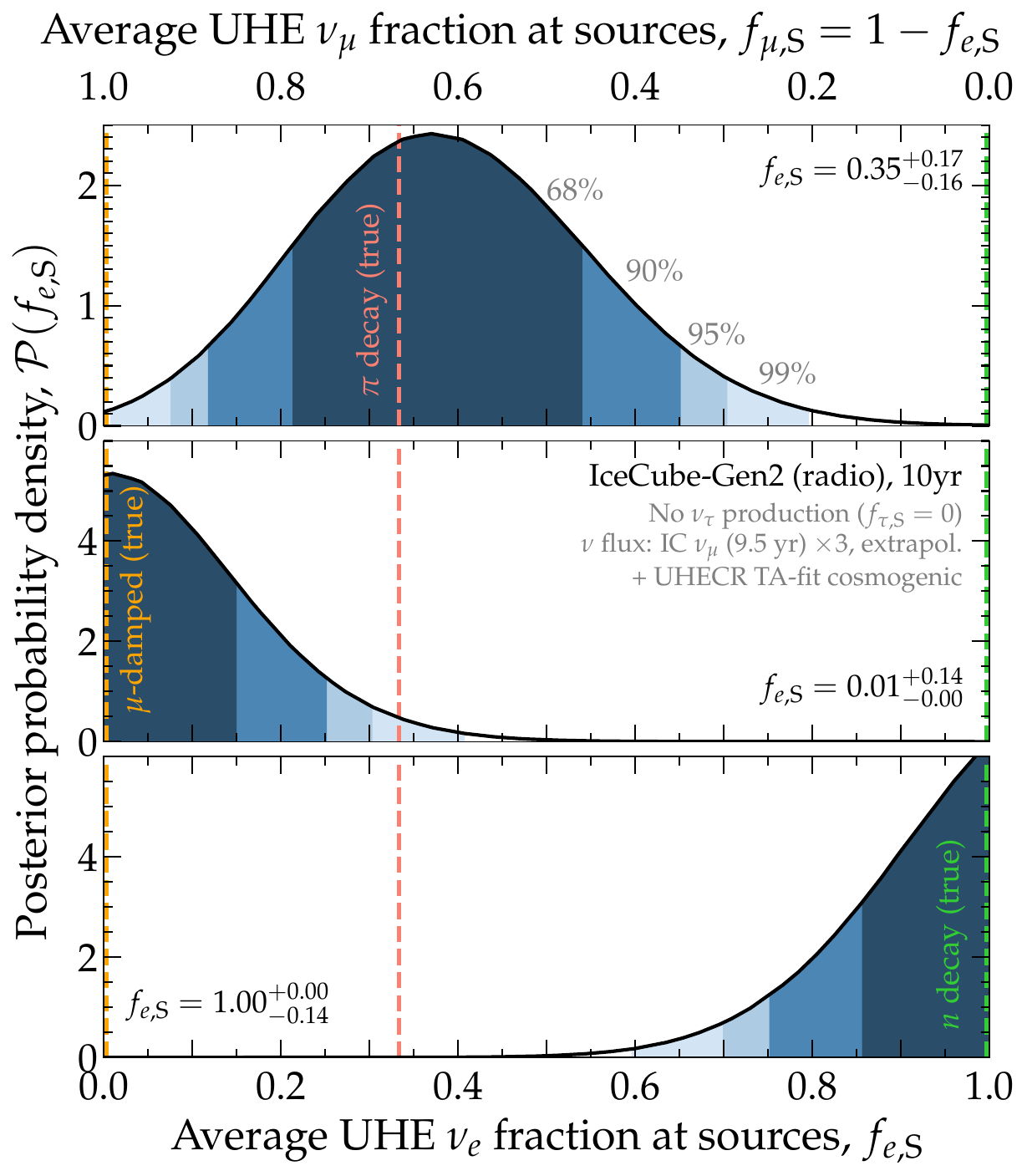}
 \caption{\textbf{\textit{Projected posterior probability distribution of the inferred fraction of UHE $\nu_e$ at their sources}}.  The true flavor composition at the sources is assumed to be the benchmark from the full pion decay ({\it top}), muon-damped pion decay ({\it center}), and neutron decay ({\it bottom}).  We assume perfect knowledge of the neutrino mixing parameters, given the expected upcoming improvement in their precision~\cite{Song:2020nfh}, and no production of $\nu_\tau$. For this figure, we assume our high benchmark UHE neutrino flux (\figu{fluxes}), derived from fitting the UHECR spectrum and mass composition to observations from TA.  See \figu{posterior_src_auger} for results assuming our low benchmark flux, \figu{posterior_src_compare} for varying detector exposure, and \Cref{sec:discussion} for details. \textit{The in-ice radio-detection of UHE neutrinos may allow us to distinguish between alternative neutrino production mechanisms that yield different flavor composition.}}
 \label{fig:posterior_src}
\end{figure}

\begin{figure}[t]
 \centering
 \includegraphics[width=\columnwidth]{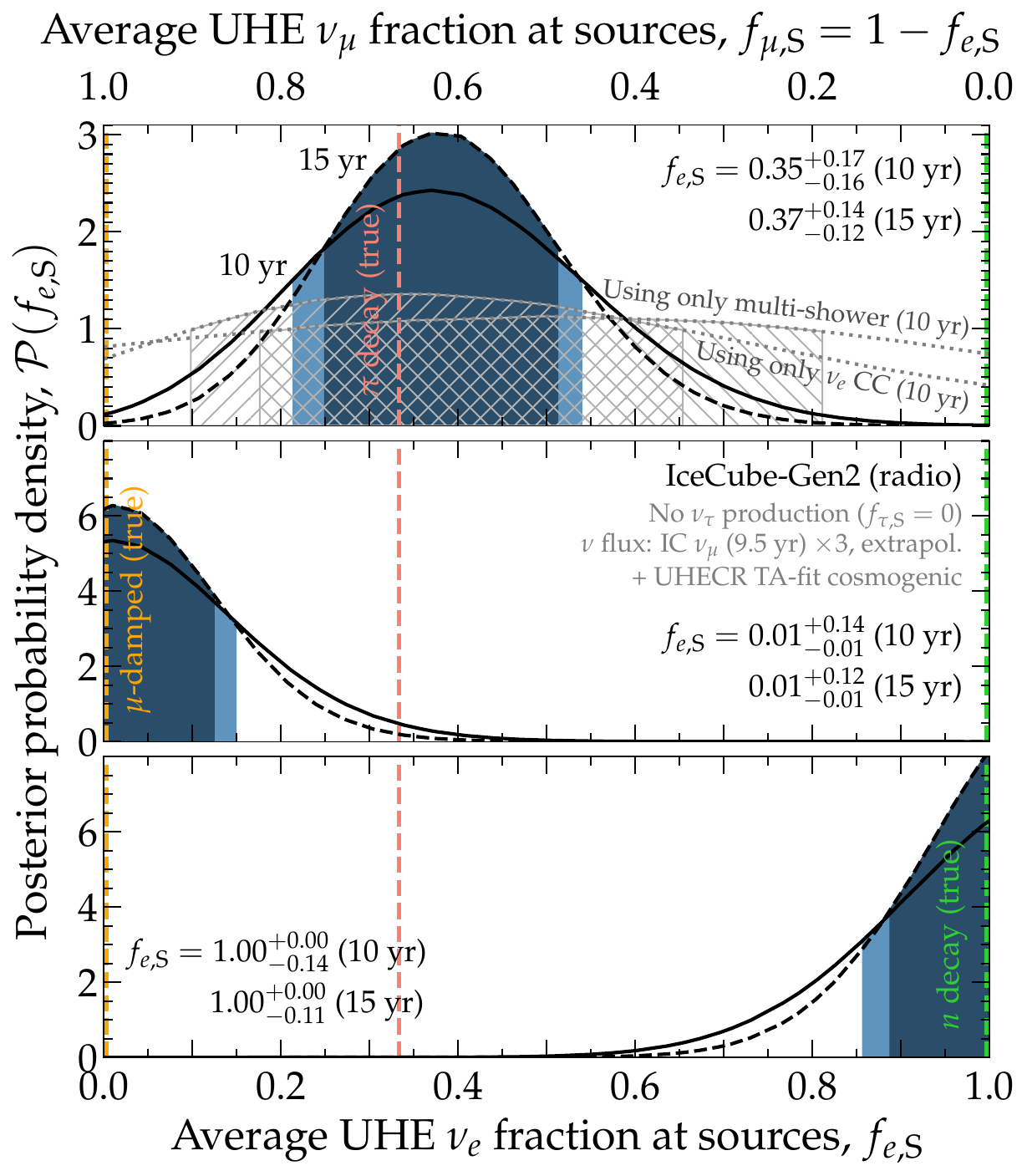}
 \caption{\textbf{\textit{Projected posterior probability distribution of the inferred fraction of UHE $\nu_e$ at their sources, varying detector exposure}}.  Similar to \figu{posterior_src}, but comparing results obtained using 10 years (as in \figu{posterior_src}) and 15 years of detector exposure.  Only for the benchmark flavor composition from the full pion decay ({\it top}), we show the result of using only one of our two flavor-sensitive channels (Sections~\ref{sec:how_to_measure_flavor-multi_shower} and \ref{sec:how_to_measure_flavor-lpm}), as in \figu{observable_pairs}.  All intervals of $f_{e, {\rm S}}$ are at 68\%~C.L.  See \figu{ternary_contours_benchmarks_ta_compare} for associated measurements of the flavor composition at Earth and \Cref{sec:discussion} for details.  \textit{Even using a single flavor-sensitive channel could provide sensitivity to the UHE $\nu_e$ at the sources.}}
 \label{fig:posterior_src_compare}
\end{figure}

Below we provide a non-exhaustive list of salient applications made available by the measurement of the UHE neutrino flavor composition, several of which extend those available by flavor measurements at lower energies~\cite{Ackermann:2019cxh, Ackermann:2019ows, Arguelles:2019rbn, Ackermann:2022rqc, Arguelles:2022tki, MammenAbraham:2022xoc}.
\begin{description}[style=unboxed]
 \item[Inferring the flavor composition at production] 
  Figure~\ref{fig:posterior_src} shows the projected flavor composition of UHE neutrinos at their sources, $f_{\alpha, {\rm S}}$, inferred from the flavor composition of UHE neutrinos measured at Earth, $f_{\alpha, \oplus}$, that we derived above, assuming our high benchmark flux.  Inferring it allows us to probe the neutrino production mechanism, the physical conditions present in the sources, and ultimately their identity.  We infer the flavor composition at the sources using the method introduced in \Refe~\cite{Bustamante:2019sdb} (see also \Refes~\cite{Song:2020nfh, Testagrossa:2023ukh}), which inverts the effect of the flavor mixing that acts during neutrino propagation.  In our projections in \figu{posterior_src}, we assume perfect knowledge of the values of the neutrino mixing parameters, which approximates the scenario expected for 2040, when the mixing angles should be known precisely enough to not add additional uncertainties~\cite{Song:2020nfh} (\Cref{sec:sensitivity_flavor}).  Depending on the true flavor composition, $f_{e, {\rm S}}$ could be inferred to within roughly 14--50\% in 10 years.  Assuming instead our low benchmark flux weakens the precision to roughly 30--80\%; see \figu{posterior_src_auger}.

  Figure~\ref{fig:posterior_src_compare} shows that increasing the detector exposure from 10 to 15 years improves the precision on $f_{e, {\rm S}}$ slightly, in agreement with the improvement on the measured $f_{\alpha, \oplus}$ shown in \figu{ternary_contours_benchmarks_ta_compare}.  Similarly to that case, exploiting the combined detection by multiple UHE neutrinos telescopes might improve the precision on $f_{e, {\rm S}}$ further; see, \eg, Fig.~2 in \Refe~\cite{Testagrossa:2023ukh}.

  Together, Figs.~\ref{fig:posterior_src} and \ref{fig:posterior_src_compare} (also \figu{posterior_src_auger}) show that the precision with which we can infer the flavor composition at the sources is better than what the underlying measurement of the flavor composition at Earth, in Figs.~\ref{fig:ternary_contours_benchmarks} and \ref{fig:ternary_contours_benchmarks_ta_compare}, might naively suggest.  This is merely because when inferring the flavor composition at the sources we only measure a single flavor fraction, $f_{e, {\rm S}}$, whereas when measuring the flavor composition at the Earth, we measure two, \eg, $f_{e, \oplus}$ and $f_{\mu, \oplus}$, since $f_{\tau, \oplus} \equiv 1- f_{e, \oplus} - f_{\mu, \oplus}$, which dilutes the precision of the measurement.
 \item[Probing the magnetic field intensity inside UHE neutrino sources] 
  Measuring a flavor composition compatible with neutrino production via muon-damped pion decay, $\left( 0,1,0 \right)_{\rm S}$, would indirectly represent a measurement of the average intensity of the magnetic field inside UHE neutrino sources that is responsible for the synchrotron-cooling of intermediate muons (\Cref{sec:sensitivity_flavor}).  Conversely, measuring a different flavor composition can be used to set upper limits on the magnetic field intensity and, by proxy, to constrain the identity of the population of the neutrino sources.  So far, this has been done using IceCube TeV--PeV neutrinos~\cite{Winter:2013cla, Bustamante:2020bxp}.  However, they provide sensitivity to  magnetic fields larger than about $10^4$~G, which are believed to be harbored only by a few candidate source populations, like a subset of gamma-ray bursts.  The synchrotron cooling of muons becomes important at energies roughly larger than $2 \cdot 10^9~(\Gamma/B)$~GeV, where $\Gamma$ is the bulk Lorentz factor of the neutrino production region and $B$ is the magnetic field intensity~\cite{Bustamante:2020bxp}.  Therefore, using UHE neutrinos would allow us to extend the sensitivity down to the 1~G scale, which would encompass significantly more candidate source populations that harbor weaker magnetic fields~\cite{Bustamante:2020bxp}.
 \item[Distinguishing cosmogenic neutrinos from UHE source neutrinos]
  Cosmogenic neutrinos are made in extragalactic space via UHE proton-photon interactions on cosmological photon backgrounds (\Cref{sec:sensitivity_flux}).  Because extragalactic magnetic fields are likely weak---nominally, of nG-scale---the flavor composition with which cosmogenic neutrinos are produced should be that of the full pion decay, \ie, $\left(\frac{1}{3},\frac{2}{3},0\right)_{\rm S}$.  Therefore, were we to measure a flavor composition compatible with muon-damped pion decay, $\left( 0, 1, 0 \right)_{\rm S}$, it would mean that the neutrinos were produced instead in the magnetized environment inside a cosmic accelerator.  Given the variety in the shape of the energy spectrum of cosmogenic neutrinos and UHE source neutrinos, measuring the UHE neutrino flavor composition could provide the only feasible way to distinguish between the two should a diffuse UHE neutrino flux be discovered.
 \item[Probing neutrino physics at the highest energies]
  Many proposed models of new physics that extend the Standard Model posit effects that grow with neutrino energy, including changes to the neutrino flavor composition.  New physics may act at neutrino production, propagation, or detection~\cite{Ackermann:2019cxh, Arguelles:2019rbn}.  Examples include, at production, nonstandard neutrino production and nonstandard neutrino-matter interactions; during propagation, pseudo-Dirac neutrinos, neutrino decay, quantum decoherence, active-sterile neutrino mixing, effective operators, dark matter-neutrino interactions, new long-range neutrino-electron and neutrino-neutron interactions, and neutrino shortcuts through extra dimensions during propagation; and, at detection, nonstandard interactions inside the Earth and the detector.  For a review, see \Refe~\cite{Rasmussen:2017ert}. 
  
  Figure~\ref{fig:ternary_he_vs_uhe} maps the regions of flavor composition at Earth accessible with standard oscillations and with two general classes of new neutrino physics, as in \Refe~\cite{Bustamante:2015waa}, but forecast using the expected uncertainty on the mixing parameters in the year 2040.  The first class resembles neutrino decay and changes the fractions of neutrino-mass eigenstates that reach Earth.  The second class resembles Lorentz-invariance violation, which modified the neutrino propagation states by augmenting the neutrino Hamiltonian with a correction whose relative importance grows with neutrino energy.  Figure~\ref{fig:ternary_he_vs_uhe} reveals that the UHE flavor sensitivity that we may achieve could potentially test extreme deviations in the flavor composition stemming from new physics, relative to the nominal expectation of approximate flavor equipartition.
  \item[TeV-to-EeV measurement with IceCube-Gen2]
   A combined measurement of the neutrino flavor with the complete IceCube-Gen2 detector allows to probe the evolution of the flavor composition with energy over six orders of magnitude from a few \si{TeV} up to \SI{10}{EeV}. The optical component of IceCube-Gen2 observes neutrinos in the TeV--PeV range~\cite{IceCube-Gen2-TDR}, whereas the radio component is sensitive to EeV neutrinos. 
   
   Figure~\ref{fig:flavour_energy} illustrates such a combined measurement. Here, we assume astrophysical sources with strong magnetic fields where the flavor composition would transition from pion decay, \ie, $\left(\frac{1}{3},\frac{2}{3},0\right)_{\rm S}$, at low energies to muon-damped production \ie, $\left(0,1,0\right)_{\rm S}$, at high energies because the decay time for secondary muons from pion decay exceeds their cooling time~\cite{Kashti:2005qa}. We assume a muon critical energy of \SI{2}{PeV} as shown in \Refe~\cite{IceCube-Gen2-TDR}. At even higher energies, the flavor composition would further transition to pion- and kaon-damped production, which will result in a steep decrease in flux, but at the same time, the flux of cosmogenic neutrinos will become dominant, leading to a transition back to a $\left(\frac{1}{3},\frac{2}{3},0\right)_{\rm S}$ flavor composition. This production model is shown as the black dashed line in \figu{flavour_energy}, together with the expected uncertainties in the flavor measurement in three energy bins. IceCube-Gen2 will be uniquely able to probe the production of cosmic neutrinos from TeV to EeV energies. 
\end{description}

\begin{figure*}[t]
 \centering
 \includegraphics[width=\textwidth]{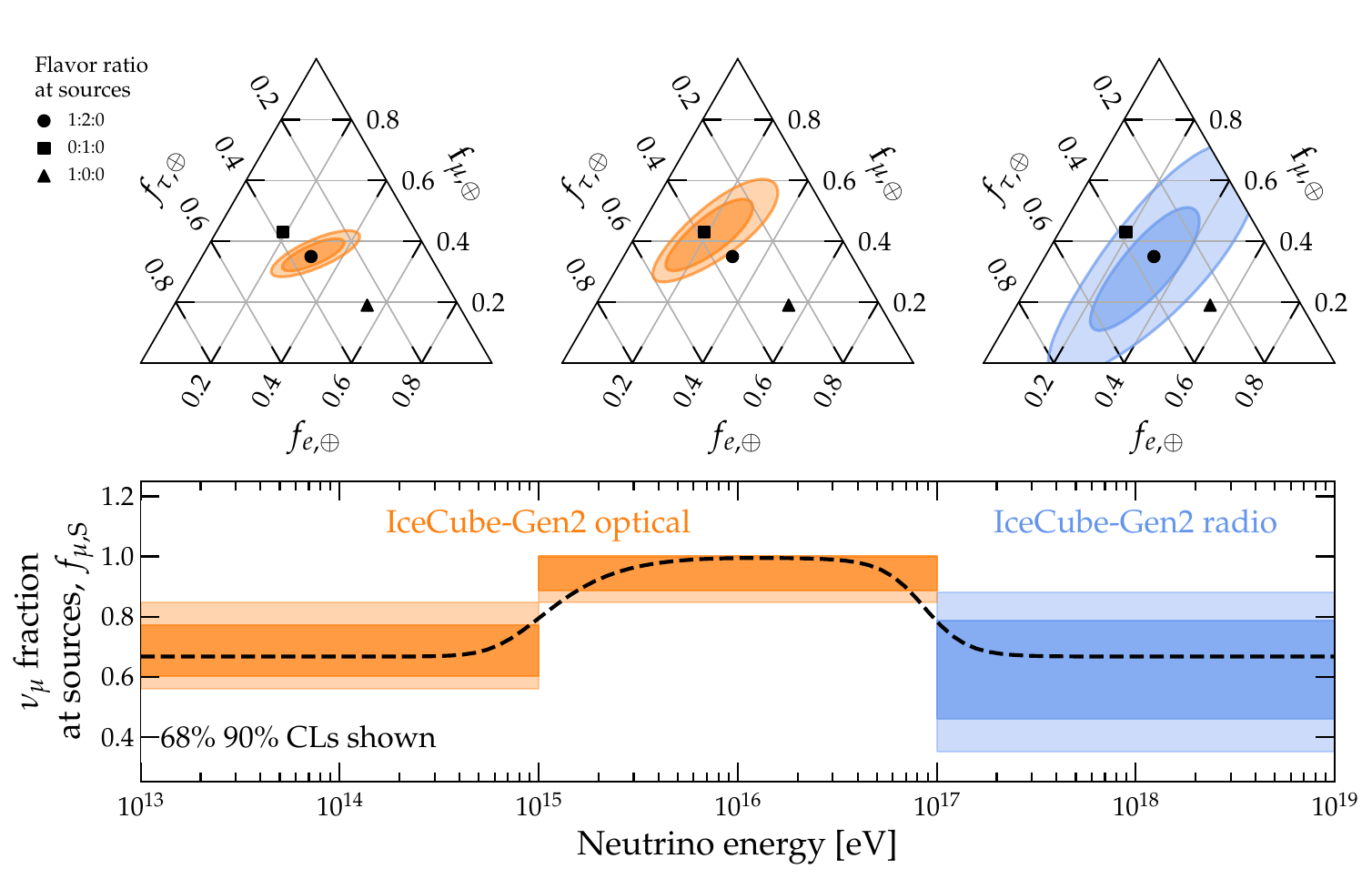}
 \caption{\textbf{\textit{Measuring the evolution of the neutrino flavor composition from TeV to EeV.}} The evolution of the neutrino flavor composition with energy can be probed by combining the radio technique described in this work with that of the optical detector of IceCube-Gen2~\cite{IceCube-Gen2-TDR}. The astrophysical neutrino sources are assumed to be dominated by pion decay at the lowest energies, $\left(\frac{1}{3},\frac{2}{3},0\right)_{\rm S}$, transition to muon-damped production at intermediate energies, $\left(0,1,0\right)_{\rm S}$, and finally return to the pion decay composition expected from cosmogenic neutrino production. The expected sensitivity of the optical component of IceCube-Gen2 was reproduced from \cite{IceCube-Gen2-TDR} using the {\sc toise} code~\cite{vanSanten:2022wss}. \textit{Top:} Uncertainty in the measurement of the flavor composition at Earth, $f_{\alpha, \oplus}$.  \textit{Bottom:} Assumed model for the evolution of $\nu_\mu$ content at the sources, $f_{\mu, {\rm S}}$, and associated uncertainty in its inferred value from measurements at IceCube. See Section~\ref{sec:discussion} for details.
    \label{fig:flavour_energy}}
\end{figure*}


\section{Summary and outlook}
\label{sec:summary}

The discovery of UHE neutrinos, with energies above $10^{17}$~eV, might finally be within reach in the near future, thanks to a new generation of large-scale neutrino telescopes that are currently under planning.  Upon discovering a diffuse flux of UHE neutrinos, measuring its flavor composition would grant us long-awaited insight into UHE neutrino physics and astrophysics to complement what can be gleamed from discovery and from measuring the neutrino energy spectrum.  Yet, in spite of its latent potential, the measurement of the UHE neutrino flavor composition has received relatively little attention.  

For the first time, in preparation for the upcoming opportunity, we have provided a method to measure the flavor composition of a diffuse UHE neutrino flux in upcoming in-ice radio-detection neutrino telescopes, such as RNO-G, presently under construction, and the radio array of IceCube-Gen2, in advanced stages of planning and to which we gear our results.  We demonstrate the proof-of-principle, but realistic feasibility of our method via state-of-the-art modeling using the {\tt NuRadioMC} code, including neutrino interactions and propagation of secondary leptons, the generation and in-ice propagation of the Askaryan radio emission they generate, and its detection by an array of underground antennas including a full detector and trigger simulation whose design matches the one envisioned for IceCube-Gen2.  

The UHE flavor sensitivity stems from two separate, complementary observables that provide access to the \nue content and to the $\numu + \nutau$ content.  On the one hand, the sensitivity to \nue comes from the identification of charged-current interactions of \nue. To identify them, we developed a dedicated neural network to identify the characteristic signature caused by the Landau-Pomeranchuk-Migdal effect.  On the other hand, the sensitivity to $\numu + \nutau$ comes from the identification of secondary interactions of muons and taus created in charged-current interactions of \numu and \nutau. In a fraction of the events, Askaryan emission from the initial neutrino interaction and a secondary interaction of the lepton can be observed in separate detector stations. 

Our results are promising. Yet, unavoidably, they depend on the size of the UHE neutrino flux, which is presently unknown.  Assuming a high diffuse UHE neutrino benchmark flux that yields about 180 detected UHE neutrinos after 10 years of observation, we find not only that it would be possible for IceCube-Gen2 to measure the flavor composition, but to achieve enough precision to distinguish between three alternative benchmarks of high-energy neutrino production that yield different flavor composition at Earth---pion decay, muon-damped pion decay, and neutron decay---at more than 68\%~C.L.  

This sensitivity, combined with envisioned precise measurements of the neutrino mixing parameters, would allow us to further enhance our ability to distinguish between production mechanisms.   Further, the measurements open up the possibility of testing extreme deviations in the flavor composition due to new neutrino physics acting at ultra-high energies.  Using a factor-of-three lower neutrino flux weakens these prospects, but does not dispel them.

Discovering UHE neutrinos is bound to bring transformative progress to the field by answering long-held extant questions in high-energy physics and astrophysics.  We have shown how to boost the insight we can gleam from them by measuring their flavor composition in mature, upcoming in-ice radio-detection neutrino telescopes.


\begin{acknowledgments}
We would like to thank the members of the IceCube-Gen2 collaboration for their useful comments and suggestions. 
CG and AC are supported by the Swedish Research Council {\sc (Vetenskapsrådet)} under project no.~2021-05449. CG is supported by the European Union (ERC, NuRadioOpt, 101116890). MB is supported by {\sc Villum Fonden} under project no.~29388.  This work used resources provided by the High-Performance Computing Center at the University of Copenhagen and resources provided by the National Academic Infrastructure for Supercomputing in Sweden (NAISS) and the Swedish National Infrastructure for Computing (SNIC) at UPPMAX partially funded by the Swedish Research Council through grant agreements no. 2022-06725 and no. 2018-05973.

\end{acknowledgments}


\appendix
\section{Additional figures}
\label{sec:appendix}

\renewcommand{\theequation}{A\arabic{equation}}
\renewcommand{\thefigure}{A\arabic{figure}}
\renewcommand{\thetable}{A\arabic{table}}
\setcounter{figure}{0} 
\setcounter{table}{0} 

\begin{figure*}
 \centering
 \includegraphics[width=0.425\textwidth]{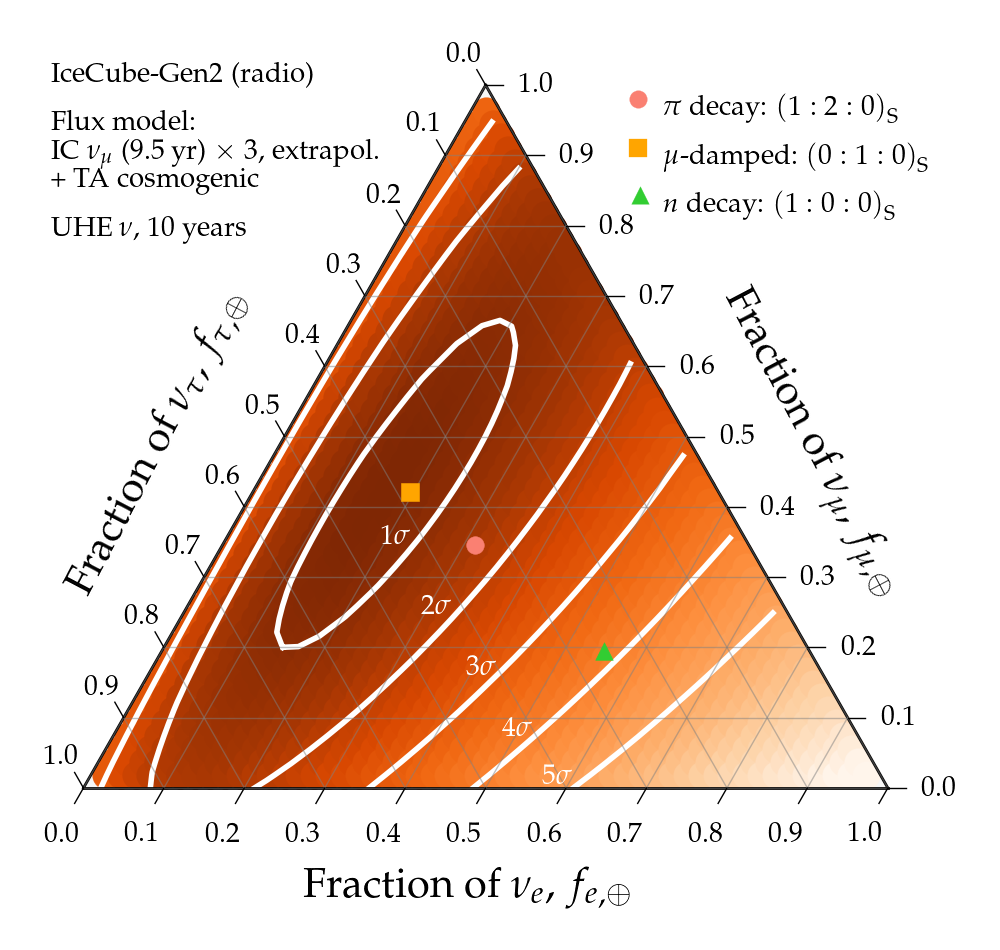}
 \hfill
 \includegraphics[width=0.425\textwidth]{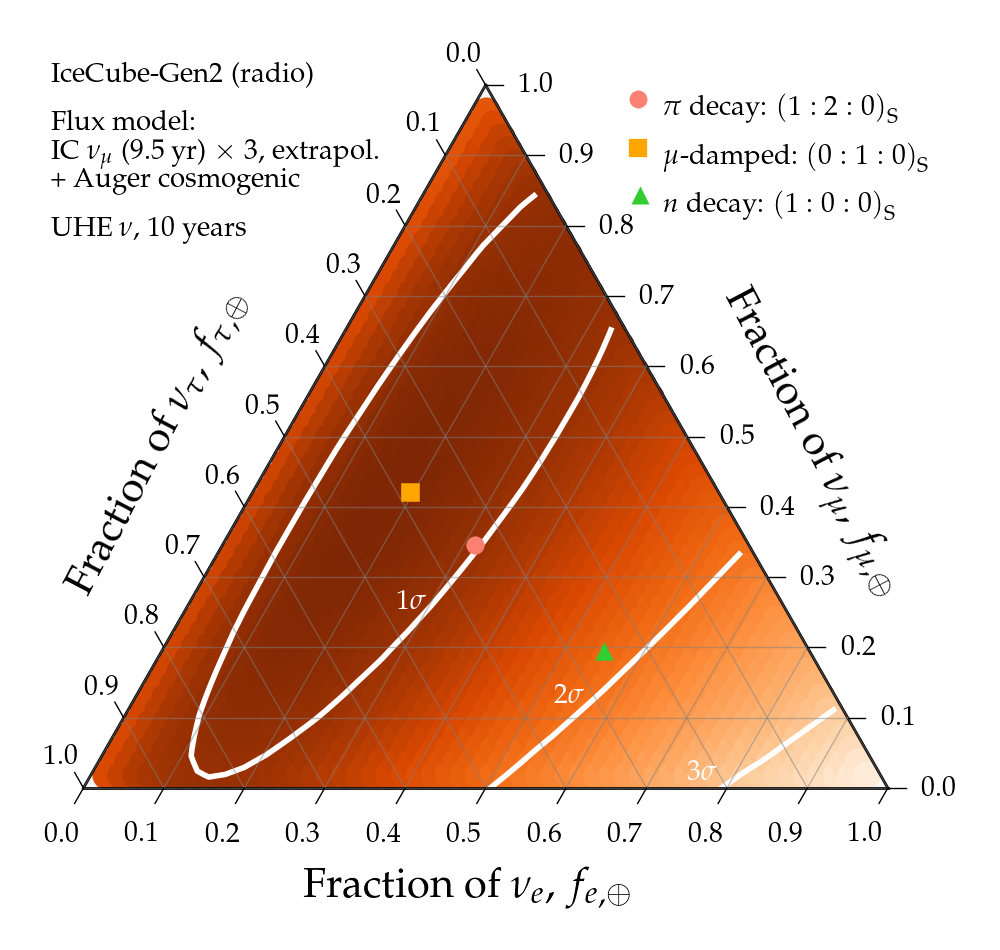}\\
 \includegraphics[width=0.425\textwidth]{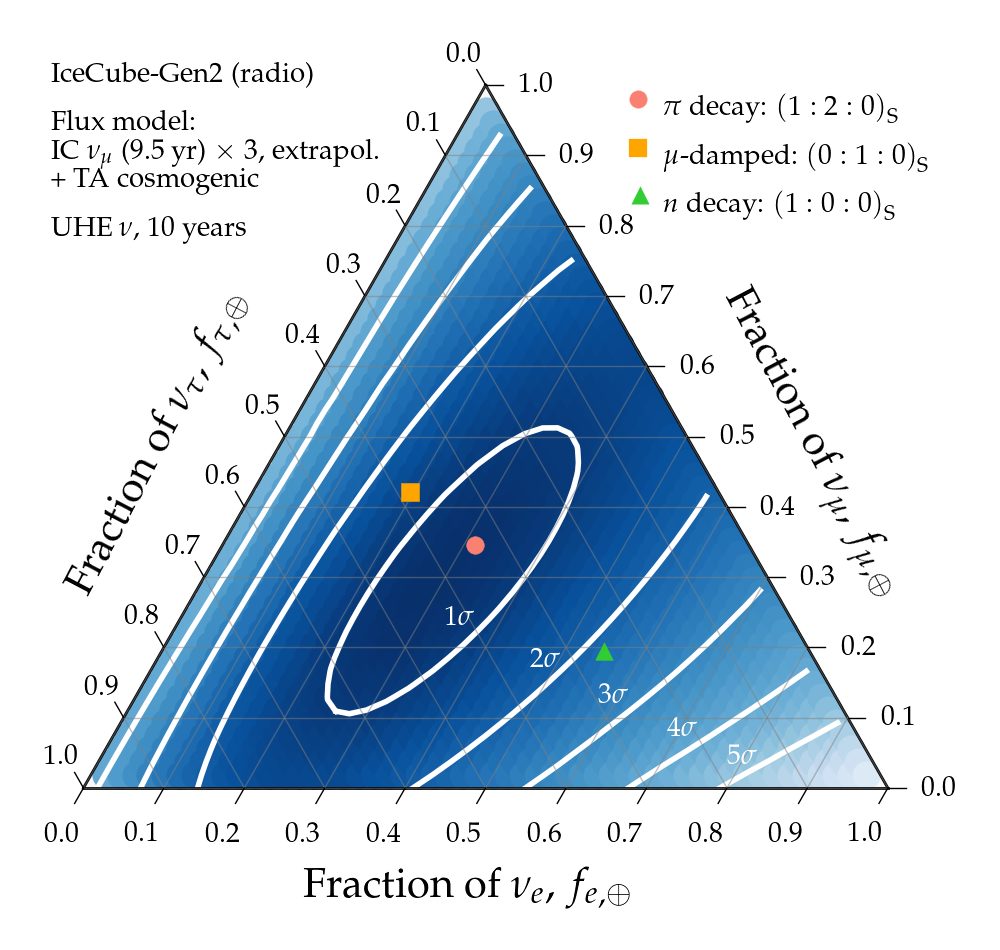}
 \hfill
 \includegraphics[width=0.425\textwidth]{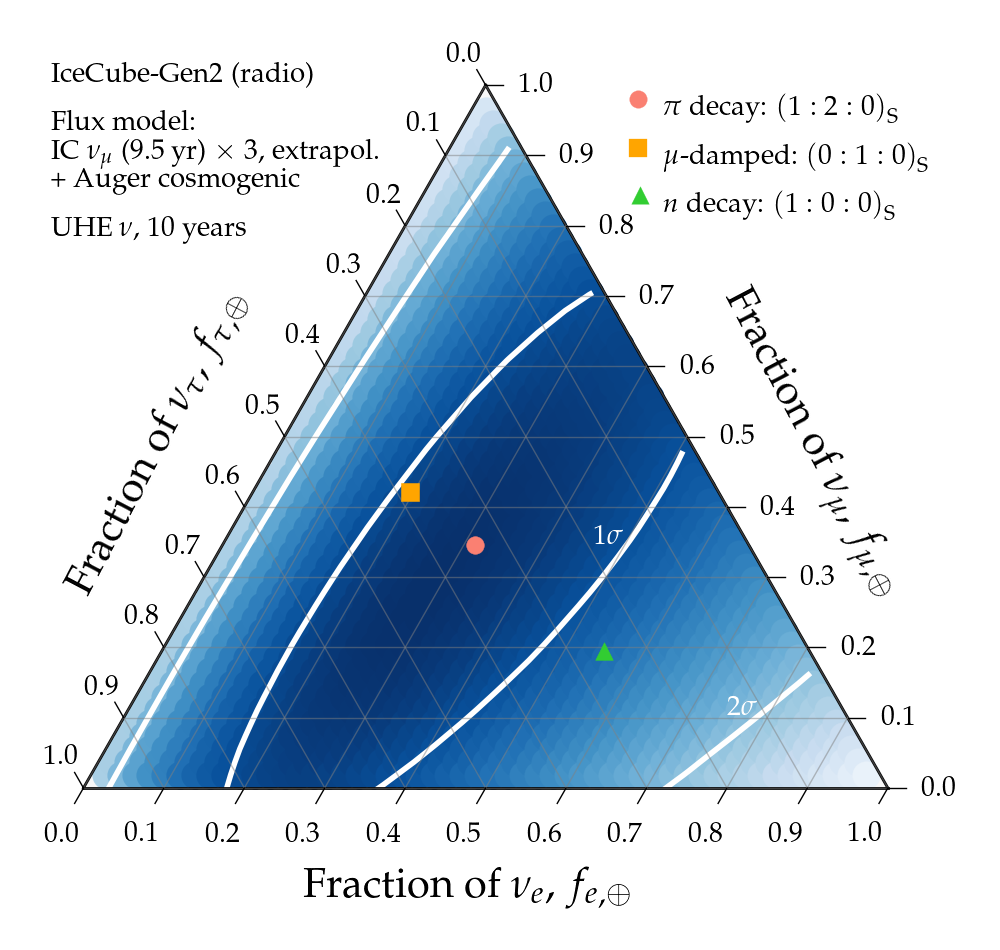}\\
 \includegraphics[width=0.425\textwidth]{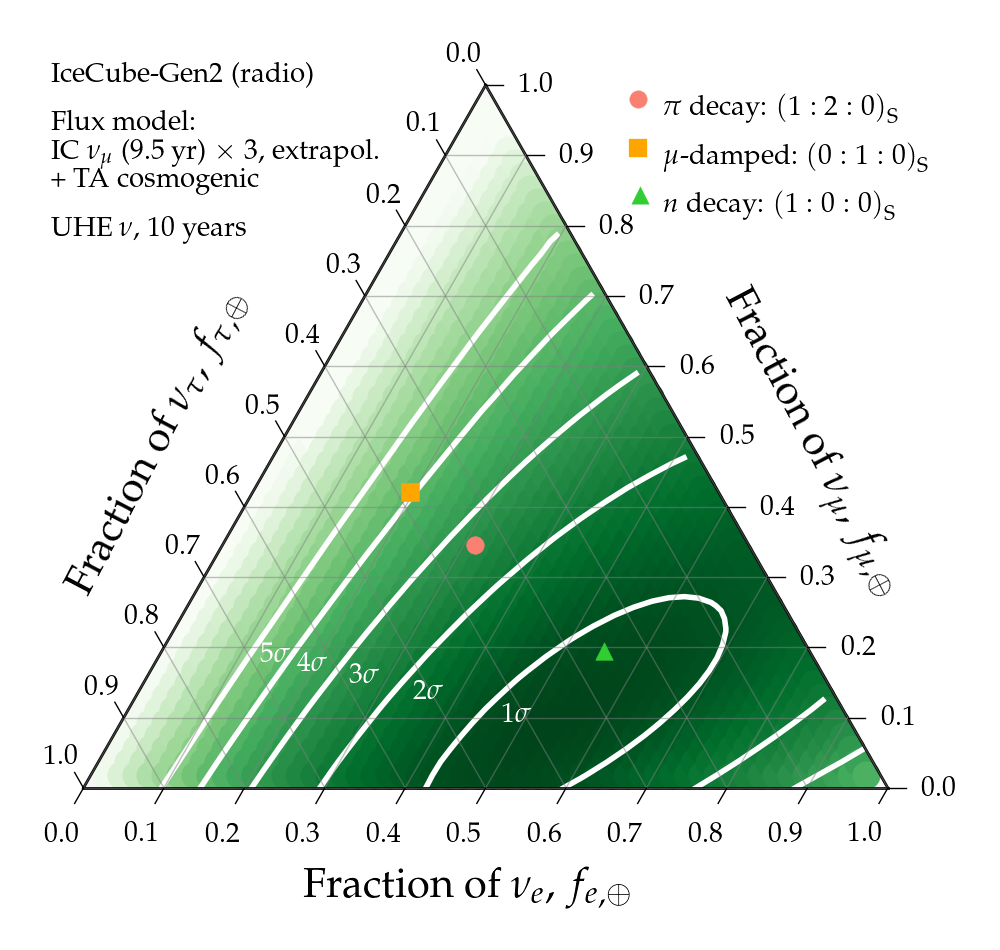}
 \hfill
 \includegraphics[width=0.425\textwidth]{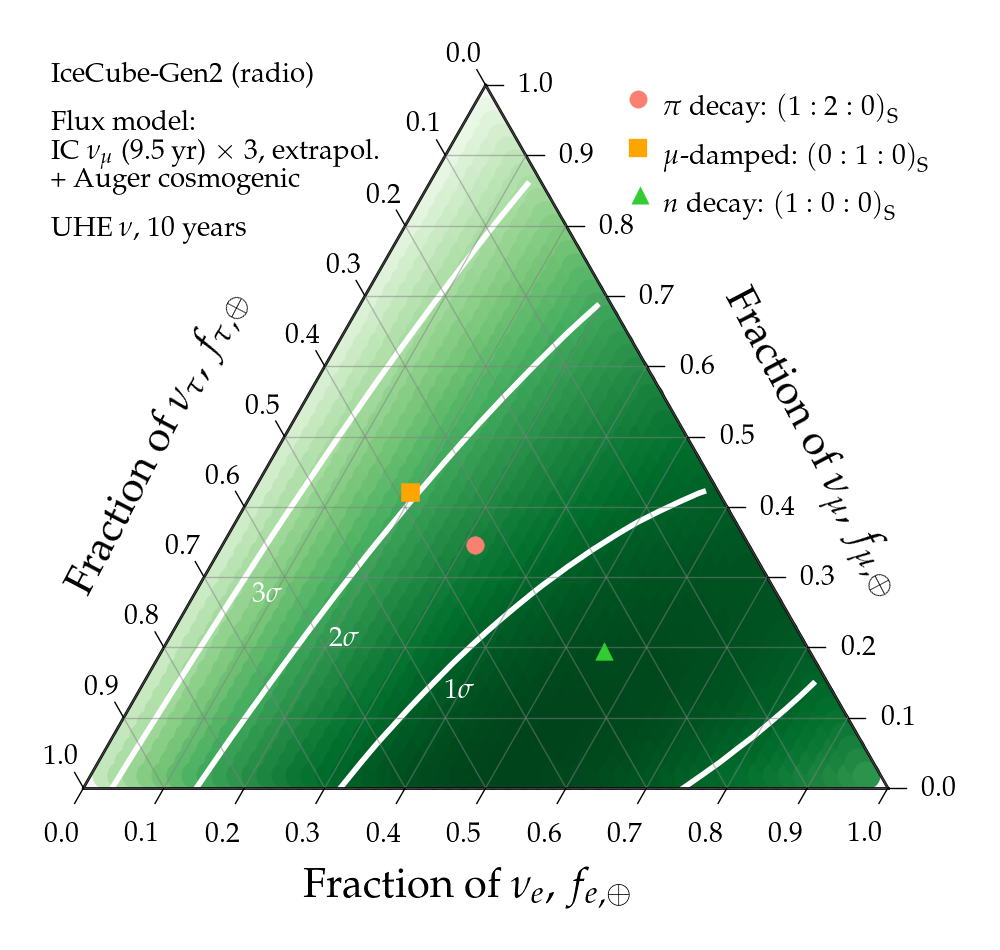}
 \caption{\textbf{\textit{Likelihood of measuring the UHE neutrino flavor composition in the radio array of IceCube-Gen2, evaluated for all possible flavor compositions at Earth.}}  The UHE neutrino flux is assumed to be our high benchmark flux ({\it left column}) or low benchmark flux ({\it right column}); see \figu{fluxes} in the main text.
 The true flavor composition at the sources is assumed to be the benchmark from the full pion decay ({\it top row}), muon-damped pion decay ({\it center row}), and neutron decay ({\it bottom row}).  The contours show allowed regions at difference confidence levels, calculated using Wilks' theorem; these are the contours shown in Figs.~\ref{fig:ternary_he_vs_uhe}, \ref{fig:observable_pairs}, and \ref{fig:ternary_contours_benchmarks}.}
 \label{fig:triangle_full}
\end{figure*}

Figure~\ref{fig:triangle_full} shows the likelihood of flavor measurement, \equ{llh_ml_mult} in the main text, evaluated at all possible combinations of the flavor composition at Earth, $(f_e, f_\mu, f_\tau)_\oplus$.  

\begin{figure}[h]
 \centering
 \includegraphics[width=\columnwidth]{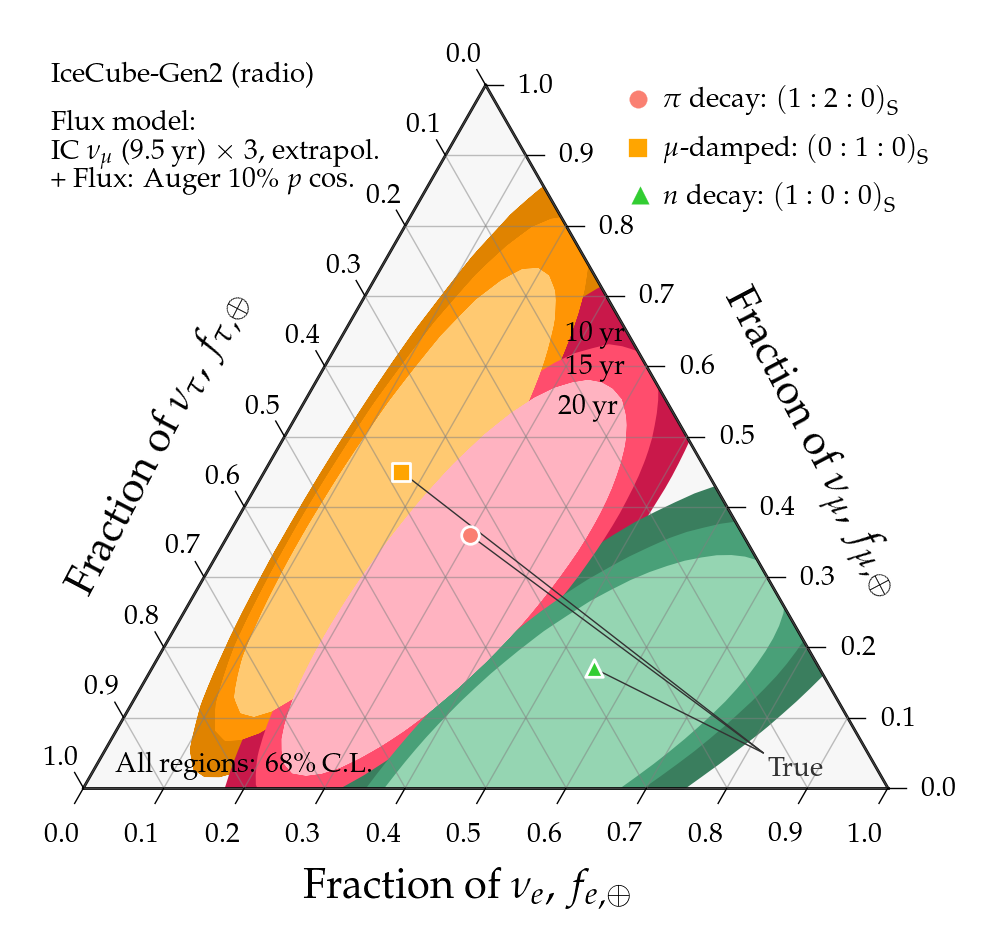}
 \caption{\textbf{\textit{Measured flavor composition of the diffuse flux of UHE neutrinos in the radio array of IceCube-Gen2,  varying detector exposure.}}  Similar to \figu{ternary_contours_benchmarks_ta_compare} in the main text, but assuming our low benchmark UHE neutrino flux (\figu{fluxes}), derived from fitting the UHECR spectrum and mass composition to observations from Auger, with 10\% of proton content. The contours correspond to 10, 15, and 20~years of data collection. See \figu{posterior_src_auger} for associated results on the inferred flavor composition at the sources.}
 \label{fig:ternary_contours_benchmarks_auger_compare}
\end{figure}

Figure~\ref{fig:ternary_contours_benchmarks_auger_compare} shows the measured flavor composition at Earth, assuming our low UHE neutrino flux benchmark, and varying the detector exposure from 10 to 20 years.  This should compared to \figu{ternary_contours_benchmarks_ta_compare} in the main text.

\begin{figure}[h]
 \centering
 \includegraphics[width=\columnwidth]{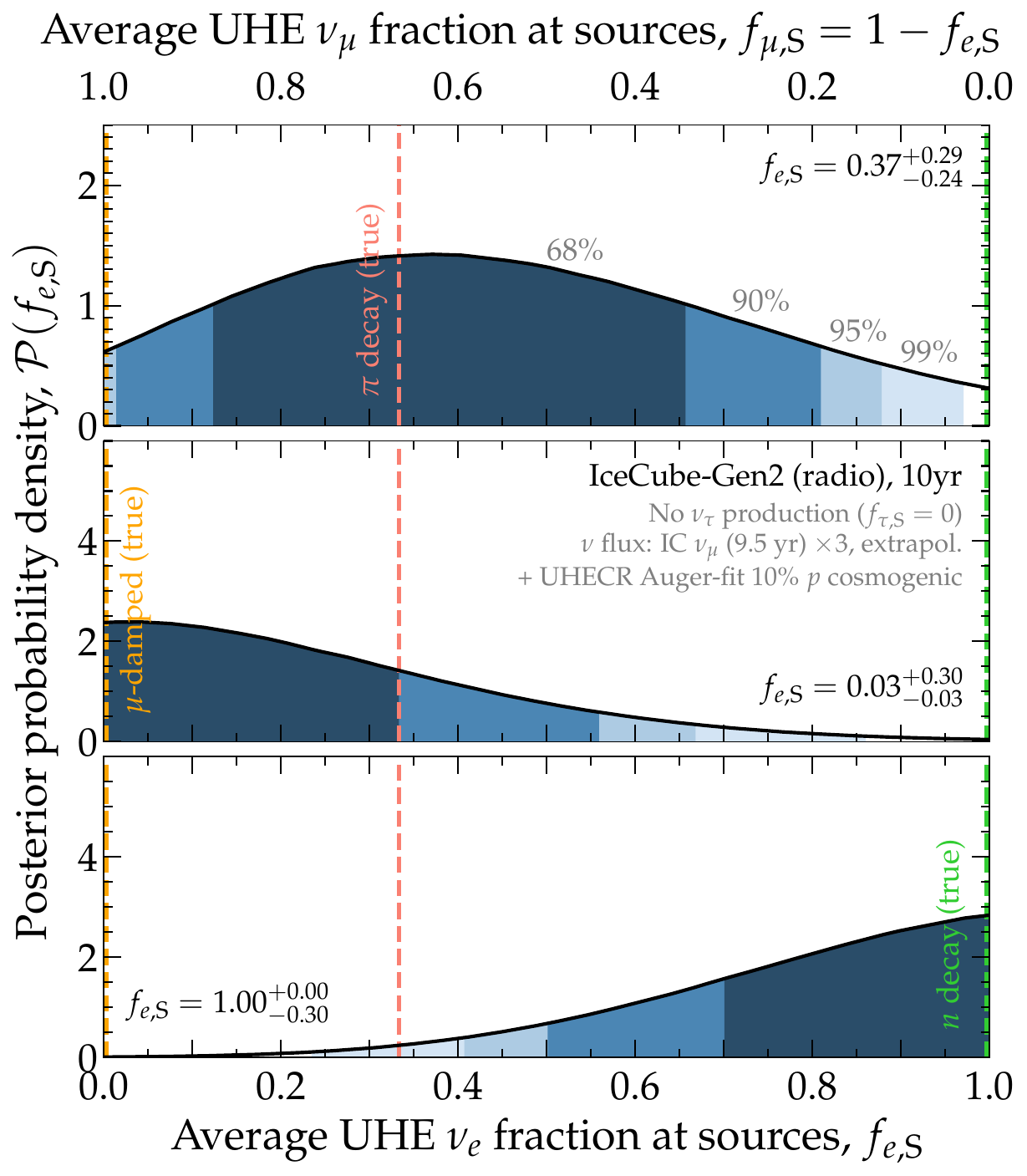}
 \caption{\textbf{\textit{Projected posterior probability distribution of the inferred fraction of UHE $\nu_e$ at their sources}}.  Same as \figu{posterior_src} in the main text, but assuming our benchmark UHE neutrino flux (\figu{fluxes}), derived from fitting the UHECR spectrum and mass composition to observations from Auger, with 10\% of proton content.  See \Cref{sec:discussion} in the main text for details.}
 \label{fig:posterior_src_auger}
\end{figure}

Figure~\ref{fig:posterior_src_auger} shows the inferred content of UHE $\nu_e$ at the sources, assuming our low UHE neutrino flux benchmark. Compared to \figu{posterior_src} in the main text, which is computed assuming our high UHE flux instead, the uncertainty in the measured value of $f_{e, \oplus}$ is about twice as large.

\bibliography{biblio.bib}

\end{document}